\documentclass[preprint2]{aastex}
\usepackage{graphicx}
\usepackage{bm}
\usepackage{amsmath}

\begin{document}
\title{POLARIZED LINE FORMATION IN MULTI-DIMENSIONAL MEDIA.III. 
HANLE EFFECT WITH PARTIAL FREQUENCY REDISTRIBUTION}
\author{L.~S.~Anusha$^{1}$ and K.~N.~Nagendra$^{1}$}
\affil{$^1$Indian Institute of Astrophysics, Koramangala,
2nd Block, Bangalore 560 034, India}

\begin{abstract}
In the previous two papers, namely, \citet{anuknn11} and \citet{anuetal11} we
solved the polarized radiative transfer (RT) equation in 
multi-dimensional (multi-D) geometries, 
with partial frequency redistribution (PRD) as the scattering mechanism.
We assumed Rayleigh scattering as the only source of linear polarization
($Q/I, U/I$) in both these papers. 
In this paper we extend these previous works to include 
the effect of weak oriented magnetic fields (Hanle effect)
on line scattering. We generalize the technique of Stokes vector
decomposition in terms of the irreducible spherical tensors
$\mathcal{T}^K_Q$, developed in \citet{anuknn11}, to the case of
RT with Hanle effect. A fast iterative method of solution (based on the 
Stabilized Preconditioned Bi-Conjugate-Gradient technique),
developed in \citet{anuetal11}, is now generalized to the case of 
RT in magnetized three-dimensional media. We use the efficient
short-characteristics formal solution method for multi-D
media, generalized appropriately to the present context.
The main results of this paper are the following:\,
(1) A comparison of emergent 
$(I, Q/I, U/I)$ profiles formed in one-dimensional (1D) media, with
the corresponding emergent, spatially averaged profiles formed in 
multi-D media, shows that in the spatially resolved structures, 
the assumption of 1D may lead to large errors in linear polarization, 
especially in the line wings.
(2) The multi-D RT in semi-infinite non-magnetic 
media causes a strong spatial variation
of the emergent $(Q/I, U/I)$ profiles, which is 
more pronounced in the line wings. 
(3) The presence of a weak
magnetic field modifies the spatial variation 
of the emergent $(Q/I, U/I)$ profiles in the line core, by producing
significant changes in their magnitudes.
\end{abstract}


\section{INTRODUCTION}
Multi-dimensional (multi-D) radiative transfer (RT) is important to advance
our understanding of the solar atmosphere. With the increase
in the resolving power of modern telescopes, and the computing power
of supercomputers, multi-D polarized line RT is becoming
a necessity, and practically feasible. The multi-D effects
manifest themselves in the resolved structures on the Sun. The finite
dimensional structures on the solar surface lead to inhomogeneity in the
atmosphere, which is then no longer axi-symmetric. The presence
of magnetic fields adds to the non-axisymmetry, in the microscopic scales
through the Hanle effect. The purpose of this paper is to address the
relative importance of non-axisymmetry caused by geometry, and
oriented magnetic fields.

In the past decades extensive studies on line RT in multi-D
media are done. A historical account on these developments is given
in \citet[][hereafter Paper I]{anuknn11}. In Paper I
we presented a method of Stokes vector decomposition, 
which helped to formulate an 
`irreducible form' of the polarized line transfer equation in a 3D 
Cartesian geometry. Such a formulation is advantageous because, the 
source vector and the mean intensity vector become angle 
independent in the reduced basis. Also the scattering phase matrix
becomes independent of the outgoing directions ($\bm{\Omega}$). 
This property leads to several advantages in numerical work.
It also provides a framework in which the transfer 
equation can be solved more conveniently, 
because the decomposition is applied to both the Stokes 
source vector, and the Stokes intensity vector. 
In \citet[][hereafter Paper II]{anuetal11}, we focused
our attention on devising fast numerical methods to solve
polarized RT equation with partial frequency redistribution (PRD)
in a two-dimensional (2D) geometry. In Paper I and Paper II 
we considered the case of non-magnetic resonance scattering
polarization. \citet{msjtb99} and \citet{ditt99} solved 
the polarized RT equation in the presence of a magnetic
field (Hanle effect), in multi-D media. Their
calculations used the assumption of complete frequency 
redistribution (CRD) in line scattering. 
In this paper we solve the same problem,
but for the more difficult and more realistic case of Hanle 
scattering with PRD. The physics of PRD scattering is
treated using the frequency-domain based approach
developed by \citet[][]{bom97a,bom97b}. The RT calculations
in one-dimensional (1D) geometry, using this approach
are described in \citet[][]{knnetal02}. We extend their
work to 2D and 3D geometries. For simplicity we restrict
to the case of angle averaged PRD functions.

The present paper represents a generalization to the magnetic
case, the decomposition technique developed in Paper I. It also
represents the generalizations to the 3D case, the Stabilized
Pre-conditioned Bi-Conjugate Gradient (pre-BiCG-STAB) method developed
in Paper II. Another generalization is the use of 
3D short characteristics formal solver in
this paper, for the case of PRD.

In Section~\ref{rte} we describe the multi-D transfer
equation in the Stokes vector basis. The decomposition technique
as applied to the case of a magnetic multi-D media
is described in Section~\ref{decomposition}. In 
Section~\ref{sc} we briefly describe the 3D 
short characteristics formal solution method. 
Section~\ref{method} is devoted to a brief description
of the numerical method of solution. Results and discussions
are presented in Section~\ref{results}. Conclusions are given
in Section~\ref{conclusions}.

\section{THE POLARIZED HANLE SCATTERING LINE TRANSFER EQUATION IN 
MULTI-D MEDIA}
\label{rte}
In this paper we consider polarized RT in 1D, 2D and 3D media
in Cartesian geometry (see Figure~\ref{fig-geometry}).
We assume that the 1D medium is infinite in the $X$ and $Y$ directions
but finite in the $Z$ direction. For 2D, we assume that the medium is 
infinite in the $X$ direction, but finite in the $Y$ and $Z$ directions.
The 3D medium is assumed to be finite in all the $X$, $Y$ and $Z$
directions. We define the ``top surface'' for a 1D medium to be the
infinite $XY$ plane passing through the point $Z_{\rm max}$.
For a 2D medium, the top surface is defined to be the plane passing through the
line $(Y, Z_{\rm max})$, which is infinite in $X$ direction.
For a 3D medium, the top surface is the plane $(X,Y, Z_{\rm max})$
which is finite in $X$ and $Y$ directions.
For a given ray with direction $\bm{\Omega}$, the 
polarized transfer equation in a multi-D medium 
with an oriented magnetic field is given by 
\begin{eqnarray}
\!\!\!\!\!\!&&\bm{\Omega} \cdot \bm{\nabla}\bm{I}(\bm{r}, \bm{\Omega}, x) = -[\kappa_l(\bm{r}) \phi(x)+\kappa_c(\bm{r})]
\nonumber \\
\!\!\!\!\!\!&& \times
[\bm{I}(\bm{r}, \bm{\Omega}, x)-\bm{S}(\bm{r}, 
\bm{\Omega}, x)],
\label{3d-rte}
\end{eqnarray}
where $\bm{I}=(I,Q,U)^T$ is the Stokes vector, with
$I$, $Q$ and $U$ the Stokes parameters defined below.
Following \citet{chandra60}, we consider an
elliptically polarized beam of light, the vibrations of the electric
vector of which describe an ellipse.
If $I_l$ and $I_r$ denote the components of the
specific intensity of this beam of light along two mutually
perpendicular directions $l$ and $r$, in a plane (see
Figure~\ref{fig-scatgeo}) transverse to the propagation 
direction, then we define
\begin{eqnarray}
I=I_l+I_r,  \nonumber \\
Q=I_l-I_r, \nonumber \\
U=(I_l-I_r) \tan 2 \chi,
\label{stokesIQU}
\end{eqnarray}
where $\chi$ is the angle between the direction $l$ and
the semi-major axis of the ellipse. Positive value of $Q$ is 
defined to be a direction parallel to $l$ and negative $Q$ 
to be in a direction parallel to $r$. The quantity 
$\bm{r}=({\rm x},{\rm y}, {\rm z})$ 
is the position vector of the ray in the Cartesian co-ordinate system.
The unit vector $\bm{\Omega}=(\eta,\gamma,\mu)=
(\sin\theta\,\cos \varphi\,,\sin\theta\,\sin \varphi\,,\cos \theta)$ 
describes the direction cosines of the ray in the atmosphere, with respect to
the atmospheric normal (the $Z$-axis), where $\theta$ and 
$\varphi$ are the polar and azimuthal angles of the 
ray (see Figure~\ref{fig-scatgeo}). The quantity $\kappa_l$ is the
frequency averaged line opacity,
$\phi$ is the Voigt profile function and $\kappa_c$ is the continuum opacity.
Frequency is measured in reduced units, namely
$x=(\nu-\nu_0)/\Delta \nu_D$, where $\Delta \nu_D$ is the Doppler width.
The Stokes source vector in a two-level atom model with unpolarized 
ground level is  
\begin{eqnarray}
&&\bm{S}(\bm{r}, \bm{\Omega}, x)\nonumber \\
&&=\frac{\kappa_l(\bm{r}) \phi(x)
\bm{S}_{l}(\bm{r}, \bm{\Omega}, x)+\kappa_c(\bm{r})
\bm{S}_{c}(\bm{r}, x)}
{\kappa_l(\bm{r}) \phi(x)+\kappa_c(\bm{r})}. \nonumber \\
\label{s-tot}
\end{eqnarray}
Here $\bm{S}_{c}$ is the continuum source vector given by
$(B_{\nu}(\bm{r}), 0, 0)^T$ with $B_{\nu}(\bm{r})$ being the 
Planck function. The line source vector is written as  
\begin{eqnarray}
&&\!\!\!\!\!\!\!\!\!\!\!\!\!\!\!\!\!\!
\bm{S}_{l}(\bm{r}, \bm{\Omega}, x)
=\bm{G}(\bm{r})+\int_{-\infty}^{+\infty}dx' \nonumber \\
&&\!\!\!\!\!\!\!\!\!\!\!\!\!\!\!\!\!\!
\times \oint\frac{d\bm{\Omega}'}{4 \pi} 
\frac{{\hat{R}}(x, x', \bm{\Omega}, \bm{\Omega}', \bm{B})} {\phi(x)} 
\bm{I}(\bm{r}, \bm{\Omega}', x'). \nonumber \\
\label{s-line}
\end{eqnarray}
Here $\hat {R}$ is the Hanle redistribution matrix and $\bm{B}$ represents
an oriented vector magnetic field. 
$\epsilon=\Gamma_I/(\Gamma_R+\Gamma_I)$
with $\Gamma_I$ and $\Gamma_R$ being the inelastic collision rate and the
radiative de-excitation rate respectively. The thermalization parameter 
$\epsilon$ is the rate of photon destruction by inelastic collisions. 
The damping parameter is computed using $a=a_R [1+(\Gamma_E+\Gamma_I)/\Gamma_R]$
where $a_R={\Gamma_R}/{4 \pi \Delta \nu_D}$ and $\Gamma_E$ is the
elastic collision rate.
We denote the thermal source vector by 
$\bm{G}(\bm{r})=\epsilon \bm{B}_{\nu}(\bm{r})$
with $\bm{B}_{\nu}(\bm{r})=(B_{\nu}(\bm{r}), 0, 0)^T$.
The solid angle
element $d \bm{\Omega}'=\sin \theta'\,d \theta'\,d 
\varphi'$, where $\theta \in [0, \pi]$ and $\varphi \in [0, 2\pi]$.
The transfer equation along the ray path takes the form
\begin{eqnarray}
&&\!\!\!\!\!\!\!\!\!\!\!\!\!\!\!\!\!\!
\frac{d\bm{I}(\bm{r}, \bm{\Omega}, x) }{ds}\nonumber \\
&&\!\!\!\!\!\!\!\!\!\!\!\!\!\!\!\!\!\!
=-\kappa_{\rm tot}(\bm{r}, x)
[\bm{I}(\bm{r}, \bm{\Omega}, x)-\bm{S}(\bm{r}, 
\bm{\Omega}, x)].\nonumber \\
\label{3d-rte-path}
\end{eqnarray}
The formal solution of Equation~(\ref{3d-rte-path}) is given by
\begin{eqnarray}
&&\!\!\!\!\!\!\!\!\bm{I}(\bm{r}, \bm{\Omega}, x)\nonumber\\
&&\!\!\!\!\!\!\!\!=\bm{I}(\bm{r}_0, \bm{\Omega}, x)
\exp\left\{-{\int_{s_0}^{s}} \kappa_{\rm tot}(\bm{r}-s^{\prime\prime}
\bm{\Omega}, x) ds^{\prime\prime}\right\}\nonumber \\
&&+\int_{s_0}^{s} \bm{S}(\bm{r}
-s^{\prime}\bm{\Omega}, \bm{\Omega}, x)
\kappa_{\rm tot}(\bm{r}-s^{\prime}\bm{\Omega},x)\nonumber\\
&&\times\exp\left\{-{\int_{s'}^{s}}  
\kappa_{\rm tot}(\bm{r}-s^{\prime\prime}
\bm{\Omega},x) ds^{\prime\prime}\right\}
ds^{\prime}.
\label{3d-formal}
\end{eqnarray}
$\bm{I}(\bm{r}_0, \bm{\Omega}, x)$ is the boundary condition
imposed at $\bm{r}_0=({\rm x}_0,{\rm y}_0,{\rm z}_0)$.
The ray path on which the formal solution is defined
is shown in Figure~\ref{fig-fs}.

\section{Decomposition of $\bm{S}$ and $\bm{I}$
for multi-D transfer in the presence of a magnetic field}
\label{decomposition}
As already discussed in Paper I, a decomposition of the Stokes source vector 
$\bm{S}$ and the intensity vector $\bm{I}$ in terms of the 
irreducible spherical tensors is necessary to simplify the problem. 
In Paper I, it was a generalization to the 3D non-magnetic case, of the
decomposition technique for the 1D transfer problems, developed by
\citet[][hereafter HF07]{hf07}. 
Here we extend our work of Paper I to include the magnetic fields.
A similar technique, but in the Fourier 
space was presented in \citet{mf91} and \citet{knnetal98}, who solved 
the Hanle scattering RT problem in 1D geometry. The solution of
polarized Hanle scattering transfer equation using the  
angle averaged and angle dependent redistribution matrices 
was presented in \citet{knnetal02}, where a perturbation method
of solution was used. A Polarized Approximate Lambda Iteration 
method to solve similar problems, using the Fourier decomposition technique
was presented in \citet{dfetal03}, but only for the case of 
angle averaged PRD.

A general theory of PRD for the 2-level atom problem with 
Hanle scattering was developed by \citet{bom97a,bom97b}.
It involves the construction of PRD matrices that describe
radiative plus collisional frequency redistribution
in scattering.
It is rather difficult to use the exact redistribution matrix $\hat {R}$ 
in the polarized transfer equation. For convenience of applications in line 
transfer theories, \citet{bom97b} proposed 3 levels of approximations, to 
handle the $\hat {R}$ matrices. In approximation levels 2 and 3, the 
$\hat {R}$ matrices were factorized into products of redistribution 
functions of \citet{hum62}, and the multi-polar components of the Hanle phase 
matrix. The collisions enter naturally in this formalism. It is
shown that such a factorization of $\hat {R}$ can be achieved only in 
certain frequency domains in the 2-dimensional $(x, x')$ frequency space. 
In this paper we refer to this way of writing the PRD Hanle $\hat {R}$ 
matrix, as the `domain based PRD'. The definition of the domains are given 
in \citet{bom97b} \citep[see also][]{knnetal02,knnetal03,dfetal03}.
We use the domain based PRD, but write the relevant equations in a
form suitable for our 
present context (multi-D transfer). We recall that in the special 
case of non-magnetic scattering, the domain based PRD equations 
for $\hat {R}$ matrix naturally go to the Domke-Hubeny 
redistribution matrix \citep{dh88}. We start by writing 
Hanle phase matrix in the atmospheric reference frame in terms of the 
irreducible spherical tensors for polarimetry, introduced
by \citet[][hereafter LL04]{ll04}. 
In this formalism the $(i,j)$-th element of the 
Hanle phase matrix is given by 
\begin{eqnarray}
&&[\hat{P}_{H}(\bm{\Omega}, \bm{\Omega'}, \bm{B})]_{ij}=\nonumber \\
&&\sum_{KQ} \mathcal{T}^K_Q(i,\bm{\Omega})\sum_{Q'}\mathcal{M}^K_{QQ'}(\bm{B})
(-1)^{Q'}\mathcal{T}^K_{-Q'}(j,\bm{\Omega'}),\nonumber \\
\end{eqnarray}
where $(i,j)=(1,2,3)$ and 
\begin{eqnarray}
&&\!\!\!\!\!\!\!
\mathcal{M}^K_{QQ'}(\bm{B})=\nonumber \\ 
&&\!\!\!\!\!\!\!e^{i(Q'-Q)\chi_B}\sum_{Q''}d^K_{QQ''}(\theta_B)
d^K_{Q''Q'}(-\theta_B) \frac{1}{1+iQ''\Gamma_B},\nonumber \\
\label{mkq}
\end{eqnarray}
where the $d^J_{MM'}$ are the reduced rotation matrices given in LL04. 
The magnetic Hanle $\Gamma_B$ parameter takes different values in 
different frequency domains (see Appendix~\ref{appendixb}). 
$\mathcal{T}^K_Q(i,\bm{\Omega})$ are the irreducible
spherical tensors for polarimetry with 
$K=0, 1, 2$, $-K \le Q \le +K$ \citep[see][]{ll04}.
In this paper, we consider only
the linear polarization. Therefore, $K=0, 2$ and $Q \in [-K, +K]$.
For the practical use, we need to further expand 
the $\hat{P}_{H}$ matrix in each of the domains in 
terms of $\mathcal{T}^K_Q$.
The required domain based expansions of the PRD matrices in
terms of $\mathcal{T}^K_Q$ were already
given in HF07, applicable there to the case of 1D Hanle transfer. We present
here the corresponding equations that are applicable to the
multi-D transfer, which
now become $\varphi$ dependent (in the 1D case, those
phase matrix components were $\varphi$ independent). 
We restrict our attention in this paper 
to the particular case of angle averaged
redistribution functions \citep[approximation level 3 of][]{bom97b}. 

The $ij$-th element of the redistribution matrix in the atmospheric 
reference frame \citep[][]{bom97b} can be written as
\begin{eqnarray}
&&\!\!\!\!\!\!\!\!\!\!\!\!\!\!\!\!
R_{ij}(x,x',\bm{\Omega},\bm{\Omega}',\bm{B})= 
\sum_{KQ} W_K \mathcal{T}^K_Q(i, \bm{\Omega}) 
\nonumber \\
&&\!\!\!\!\!\!\!\!\!\!\!\!\!\!\!\!\!\!\!\!\times \Big \{ r_{\rm II}(x, x') 
P^K_{Q, \rm II}(j,\bm{\Omega}',\bm{B}) \nonumber \\
&&+r_{\rm III}(x, x') 
P^K_{Q, \rm III}(j,\bm{\Omega}',\bm{B})\Big \}.
\label{redist-tkq}
\end{eqnarray}
The weights $W_K$ depend on the line under consideration (see LL04).
Here $r_{\rm II}(x, x')$ and $r_{\rm III}(x, x')$ are the angle-averaged
versions of redistribution functions \citep[see][]{hum62}.
The quantities $P^K_{Q, \rm II}(j,\bm{\Omega}',\bm{B})$ and 
$P^K_{Q, \rm III}(j,\bm{\Omega}',\bm{B})$ take 
different forms in different frequency domains. They are described in
Appendix~\ref{appendixb}.

Denoting $G^K_Q = \delta_{K0} \delta_{Q0} G(\bm{r})$, where
$G(\bm{r})= \epsilon B_{\nu}(\bm{r})$, we can write the $i$-th 
component of the thermal source vector as
\begin{equation}
G_{i}(\bm{r})=\sum_{KQ}\mathcal{T}^K_Q(i, \bm{\Omega})G^K_Q(\bm{r}).
\label{gkq}
\end{equation}
The line source vector can be decomposed as
\begin{equation}
S_{i, l}(\bm{r}, \bm{\Omega}, x)=\sum_{KQ}
\mathcal{T}^K_Q(i, \bm{\Omega})S^K_{Q,l}(\bm{r}, x),
\label{skq-decompose}
\end{equation}
where
\begin{eqnarray}
&&\!\!\!\!\!\!\!\!\!\!\!
S^K_{Q,l}(\bm{r}, x)=G^K_Q(\bm{r})+\frac{1}{\phi(x)}
\int_{-\infty}^{+\infty}dx' \oint
\frac{d \bm{\Omega}'}{4 \pi}\nonumber \\
&&\!\!\!\!\!\!\!\!\!\!\!
\times \sum_{j=0}^{3}
W_K \Big\{r_{\rm II}(x, x')
{P}^K_{Q, \rm II}(j, \bm{\Omega}',\bm{B})
\nonumber \\
&&\!\!\!\!\!\!
+r_{\rm III}(x, x')
{P}^K_{Q, \rm III}(j, \bm{\Omega}',\bm{B})\Big\}
I_{j}(\bm{r}, \bm{\Omega}', x'). \nonumber \\
\label{skq}
\end{eqnarray}
Note that the components $S^K_{Q,l}(\bm{r}, x)$ now depend only on the
spatial variables $({\rm x}, {\rm y}, {\rm z})$, 
frequency $x$.
The $(\theta, \varphi)$ dependence is fully contained in 
$\mathcal{T}^K_{Q}(i, \bm{\Omega})$. These quantities are listed in LL04
(chapter 5, Table 5.6, p.~211).
Substituting Equation~(\ref{skq-decompose}) in
Equation~(\ref{3d-formal}),
the components of $\bm{I}$ can be written as
\begin{eqnarray}
\!\!\!\!\!\!\!\!\!&&I_{i}(\bm{r}, \bm{\Omega}, x)=
\sum_{KQ}\mathcal{T}^K_Q(i, \bm{\Omega}) {I}^K_{Q}(\bm{r},\bm{\Omega}, x),
\nonumber \\
\label{ikq-decompose}
\end{eqnarray}
where
\begin{eqnarray}
&&\!\!\!\!\!\!\!\!\!{I}^K_{Q}(\bm{r},\bm{\Omega}, x)=
{I}^K_{Q, 0}(\bm{r}_0, \bm{\Omega}, x)
e^{-\tau_{x, {\rm max}}}\nonumber \\
&&\!\!\!\!\!\!\!\!\!
+\int_{0}^{\tau_{x, {\rm max}}} 
e^{-\tau'_{x}(\bm{r}')}
\Big[p_x S^K_{Q, l}(\bm{r}', x) \nonumber \\
&&\!\!\!\!\!\!\!\!\!+ (1-p_x) S^K_{Q, C}(\bm{r}', x) 
\Big]\,d\tau'_{x}(\bm{r}').
\nonumber \\
\label{ikq-out-tau}
\end{eqnarray}
Here $I^K_{Q,0}=I_0(\bm{r}_0, \bm{\Omega}, x) \delta_{K0} \delta_{Q0}$
are the intensity
components at the lower boundary. The quantities
$S^K_{Q, C}=S_C(\bm{r}, x) \delta_{K0}  \delta_{Q0}$ denote
the continuum source vector
components. We assume that $S_C(\bm{r}, x)=B_{\nu}(\bm{r})$. The ratio of the
line opacity to the total opacity is given by
\begin{equation}
p_x=\kappa_l(\bm{r}) \phi(x) / \kappa_{\rm tot}(\bm{r}, x).
\label{px}
\end{equation}
The monochromatic optical depth scale is defined as
\begin{equation}
\tau_x({\rm x},{\rm y}, {\rm z})=\int_{s_0}^s \kappa_{\rm tot}(\bm{r}-s''
\bm{\Omega}, x)\, ds'',
\label{tau}
\end{equation}
where $\tau_x$ is measured along
a given ray determined by the direction $\bm{\Omega}$.
In Equation~(\ref{ikq-out-tau}) $\tau_{x, {\rm max}}$ is the maximum
monochromatic optical depth at frequency $x$, when measured along the ray.
\subsection{The irreducible transfer equation in multi-D geometry for the Hanle 
scattering problem}
\noindent
Let $S^K_Q=p_xS^K_{Q, l}+(1-p_x)S^K_{Q, C}$.
$I^K_Q$ and $S^K_Q$ as well as the phase matrix elements 
$P^K_{Q, \rm II}(j,\bm{\Omega}',\bm{B})$ and
$P^K_{Q, \rm III}(j,\bm{\Omega}',\bm{B})$
are all complex quantities. 
Following the method of transformation from complex
to the real quantities given in HF07, we define the real
irreducible Stokes vector $\bm{\mathcal I}=(I^0_0$, 
$I^2_0$, $I^{2,{\rm x}}_1$, $I^{2,{\rm y}}_1$, 
$I^{2,{\rm x}}_2$, $I^{2,{\rm y}}_2)^T$ and the
real irreducible source vector $\bm{\mathcal S}=(S^0_0$, 
$S^2_0$, $S^{2,{\rm x}}_1$, $S^{2,{\rm y}}_1$, 
$S^{2,{\rm x}}_2$, $S^{2,{\rm y}}_2)^T$.
It can be shown that the 
$\bm{\mathcal I}$ and $\bm{\mathcal S}$ satisfy a transfer
equation of the form
\begin{eqnarray}
&&-\frac{1}{\kappa_{\rm tot}(\bm{r}, x)}\bm{\Omega} \cdot
\bm{\nabla}\bm{\mathcal{I}}(\bm{r}, \bm{\Omega}, x) = \nonumber \\
\!\!\!\!\!\!&&[\bm{\mathcal{I}}(\bm{r},\bm{\Omega}, x)-
\bm{\mathcal S}(\bm{r}, x)],
\label{rte-real-reduced}
\end{eqnarray}
where $\bm{\mathcal S}(\bm{r}, x)=p_x 
\bm{\mathcal S}_{l}(\bm{r}, x)
+(1-p_x)\bm{\mathcal S}_C(\bm{r}, x)$ with
\begin{eqnarray}
\!\!\!\!\!\!&&\bm{\mathcal S}_{l}(\bm{r}, x)=
\epsilon \bm{B}_{\nu}(\bm{r})\nonumber \\
\!\!\!\!\!\!&&+\frac{1}{\phi(x)} \int_{-\infty}^{+\infty} 
dx' 
\oint\frac{d \bm{\Omega}'} {4 \pi} 
\hat{W}\Big\{\hat{M}^{(i)}_{\rm II}(\bm{B}){r}_{\rm II}(x, x') \nonumber \\
\!\!\!\!\!\!&&+\hat{M}^{(i)}_{\rm III}(\bm{B})
{r}_{\rm III}(x, x') \Big\}
\hat{\Psi}(\bm{\Omega}') 
\bm{\mathcal{I}}(\bm{r}, \bm{\Omega}',x'),\nonumber \\
\label{sl-real-reduced}
\end{eqnarray}
and $\bm{\mathcal S}_C(\bm{r}, x)$=$(S_C(\bm{r},x),0,0,0,0,0)^T$.
$\hat{W}$ is a diagonal matrix given by
\begin{equation}
\hat{W}=\textrm{diag}\{W_0,W_2,W_2,W_2,W_2,W_2\}.
\label{w}
\end{equation}
The matrix $\hat{\Psi}$ represents the phase matrix for
the Rayleigh scattering, to be used in multi-D geometries. Its 
elements are listed in Appendix~\ref{appendixd}.
The matrices $\hat{M}^{(i)}_{\rm II,III}(\bm{B})$ in
different domains are given in Appendix~\ref{appendixc}.
The formal solution now takes the form
\begin{eqnarray}
&&\!\!\!\!\!\!\!\!\!\bm{\mathcal{I}}(\bm{r},\bm{\Omega}, x)=
\bm{\mathcal {I}}(\bm{r}_0, \bm{\Omega}, x)
e^{-\tau_{x, {\rm max}}}\nonumber \\
&&\!\!\!\!\!\!\!\!\!
+\int_{0}^{\tau_{x, {\rm max}}} 
e^{-\tau'_{x}(\bm{r}')}
\bm{\mathcal {S}}(\bm{r}', x)d\tau'_{x}(\bm{r}').
\nonumber \\
\label{i-real-out-tau}
\end{eqnarray}
Here $\bm{\mathcal {I}}(\bm{r}_0, \bm{\Omega}, x)$ 
is the boundary condition imposed at $\bm{r}_0$.
\section{A 3D FORMAL SOLVER BASED ON THE SHORT CHARACTERISTICS APPROACH}
\label{sc}
This section is devoted to a discussion of 3D short characteristics 
formal solver. Here we generalize to the 3D case, the 2D short characteristics
formal solver that we had used in Paper II. 
A short characteristic stencil ${\rm MOP}$ of a ray
passing through the point $\rm O$, in a 3D cube is shown in 
Figure~\ref{fig-sc}. The point $\rm O$ represents a grid point along the
ray path. The point $\rm M$ (or $\rm P$) represents an 
intersection of the ray with one of the boundary
planes of a 3D cell. The plane of intersection is determined by
the direction cosines of the ray. The length $\Delta s$ of the line
segment ${\rm MO}$ (or ${\rm OP}$) is given by
\begin{eqnarray}
&&\Delta s = \Delta {\rm z}/ \mu, 
\quad{\textrm {if the ray hits the $XY$ plane}},\nonumber\\
&&\Delta s = \Delta {\rm y}/ \gamma, 
\quad{\textrm {if the ray hits the $XZ$ plane}},\nonumber\\
&&\Delta s = \Delta {\rm x}/ \eta, 
\quad{\textrm {if the ray hits the $YZ$ plane}}.\nonumber\\
\label{deltaxyz}
\end{eqnarray}
Here $\Delta {\rm x}$, $\Delta {\rm y}$ and $\Delta {\rm z}$
are incremental lengths (positive or negative) between two successive
grid points on the $X$, $Y$ and $Z$ directions respectively.
In the short characteristics method, the irreducible Stokes vector
$\bm{\mathcal{I}}$ at ${\rm O}$ is given by
\begin{eqnarray}
&&\bm{\mathcal{I}}_{\rm O}(\bm{r}, \bm{\Omega}, x)=
\bm{\mathcal{I}}_{\rm M}(\bm{r}, \bm{\Omega}, x) 
\exp[-\Delta \tau_{\rm M}]
\nonumber \\
&&+{\psi}_{\rm M}(\bm{r}, \bm{\Omega}, x)
\bm{\mathcal{S}}_{\rm M}(\bm{r}, x)\nonumber \\
&&+{\psi}_{\rm O}(\bm{r}, \bm{\Omega}, x)
\bm{\mathcal{S}}_{\rm O}(\bm{r}, x)\nonumber \\
&&+{\psi}_{\rm P}(\bm{r}, \bm{\Omega}, x)
\bm{\mathcal{S}}_{\rm P}(\bm{r}, x),\nonumber \\ 
\label{int_sc}
\end{eqnarray}
where $\bm{\mathcal{S}}_{\rm M,O,P}$ are the irreducible source vectors at
${\rm M}$, ${\rm O}$ and ${\rm P}$.
The quantity $\bm{\mathcal{I}}_{\rm M}$ is the upwind irreducible Stokes
vector for the point ${\rm O}$. If ${\rm M}$ and ${\rm P}$ are non-grid points,
then $\bm{\mathcal{S}}_{\rm M,P}$ and $\bm{\mathcal{I}}_{\rm M}$ are
computed using a two-dimensional parabolic interpolation formula.
While computing them, one has to ensure the monotonicity of all the 6
components of these vectors, through appropriate logical tests
\citep[see][]{auerfp94}.
The coefficients ${\psi}$ depend on the optical depth increments in 
$X$, $Y$ and $Z$ directions. For a 2D geometry, these coefficients
are given in \citet{auerfp94}. Here we have used a generalized version
of these coefficients, that are applicable to a 3D geometry.

\section{NUMERICAL METHOD OF SOLUTION}
\label{method}
In this paper we generalize the pre-BiCG-STAB method described in Paper II 
to the case of a 3D geometry.
The present work represents also an extension of this technique 
to the case of polarized
RT in the presence of an oriented magnetic field. The essential difference
between the 2D and 3D algorithms is in terms of the lengths of the vectors.
In a 2D geometry it is $n_p\times n_x\times n_Y\times n_Z$ 
whereas in a 3D geometry it is
$n_p\times n_x \times n_X \times n_Y\times n_Z$, where $n_{X,Y,Z}$ are
the number of grid points in the $X$, $Y$ and $Z$ directions, and $n_x$ refers
to the number of frequency points. $n_p$ is the number of polarization 
components of the irreducible vectors. In the presence of a magnetic
field, $n_p=6$ in both 2D and 3D geometries. In non-magnetic problems,
$n_p=4, 6$ for 2D and 3D geometries respectively.
\subsection{The Preconditioner matrix}
\label{preconditioner}
A description of the preconditioner matrix that appears in the
pre-BiCG-STAB method, is already given in Paper II. Here we give
its functional form applicable to the problems considered in this paper.
In Paper II a single preconditioner matrix was sufficient to handle
the non-magnetic line transfer problem with PRD. The presence of
magnetic field requires the use of domain based PRD matrices, for
a better description of the PRD in line scattering. The method
requires preconditioner matrices to be defined, that are suitable
for each of the frequency domains. We denote the preconditioner
matrices by $\hat{\mathcal {M}}^{(i)}$.
\begin{eqnarray}
&&\hat{\mathcal {M}}^{(i)}=\hat{I}-p_x\nonumber\\
&&\times\frac{1}{\phi(x)}\Big\{\Lambda^{\star\,(i)}_{x',\rm II}
{r}_{\rm II}(x,x')+\Lambda^{\star\,(i)}_{x',\rm III})
{r}_{\rm III}(x,x')\Big\},\nonumber\\
\label{Minv}
\end{eqnarray}
where
\begin{eqnarray}
\Lambda^{\star\,(i)}_{x',\rm II}=
\oint \frac{d \bm{\Omega}'} {4 \pi} 
\hat{W}\hat{M}^{(i)}_{\rm II}(\bm{B})
%\nonumber \\
%\times 
\hat{\Psi}(\bm{\Omega}') 
\bm{\mathcal{I}}(\bm{r}, \bm{\Omega}',x'),
\label{lambdastar1}
\end{eqnarray}
and
\begin{eqnarray}
\Lambda^{\star\,(i)}_{x',\rm III}=
\oint \frac{d \bm{\Omega}'} {4 \pi} 
\hat{W}\hat{M}^{(i)}_{\rm III}(\bm{B})
%\nonumber \\
%\times 
\hat{\Psi}(\bm{\Omega}') \bm{\mathcal{I}}(\bm{r}, \bm{\Omega}',x').
\label{lambdastar2}
\end{eqnarray}
Here $\bm{\mathcal{I}}(\bm{r},\bm{\Omega}',x')$ 
is computed using a delta source vector as input.
The expressions for the matrices $\hat{M}^{(i)}_{\rm II}$ 
and $\hat{M}^{(i)}_{\rm III}$ 
in different domains are given in Appendix~\ref{appendixc}.
The matrices $\hat{\mathcal{M}}^{(i)}$ are block diagonal. 
Each block is a full matrix with respect to $x$ and $x'$. 
The matrices $\hat{\mathcal{M}}^{(i)}$ are 
diagonal with respect to other variables.

\subsection{Computational details}
\label{comp-detail}
To calculate the integral in Equation~(\ref{sl-real-reduced}) and the
formal solution in Equation~(\ref{int_sc}), we need to define
quadratures for angles, frequencies and depths.

For all the computations presented in this paper, Carlsson type B 
angular quadrature with an order $n=8$ is used. 
All the results are presented in this paper for damping parameter 
$a=10^{-3}$. The number of frequency points required for a given problem 
depends on the value of $a$ and the optical thickness in the $X$, $Y$ 
and $Z$ directions (denoted by $T_X$, $T_Y$ and $T_Z$). A frequency 
bandwidth satisfying the conditions $\phi(x_{\rm max})T_X < < 1$, 
$\phi(x_{\rm max})T_Y < < 1$ and $\phi(x_{\rm max})T_Z < < 1$ at the 
largest frequency point denoted by $x_{\rm max}$ has been used.
We have used a logarithmic frequency grid with a fine spacing in the
line core region, and the near wings where the PRD effects are important.
We use a logarithmic spacing in the $X$, $Y$ and $Z$ directions, with a 
fine griding near the boundaries. We find that with the modern solution
methods used in the calculations give sufficiently accurate solutions 
for 5 spatial points per decade.

Computing time depends on the number of angle, frequency and depth
points considered in the calculations and also the machine used
for computations. We use the Intel(R) Core(TM) i5 CPU 760 at 
2.8 GHz processor running an un-parallelized code. For the 
difficult test case of a semi-infinite 3D atmosphere the computing time
is approximately an hour for one iteration. Even for this difficult
test case the Pre-BiCG-STAB method needs just 18 iterations 
to reach a convergence criteria of $10^{-8}$.

\section{RESULTS AND DISCUSSIONS}
\label{results}
In this section we present the results of computations to illustrate 
broader aspects of the polarized transfer in 1D, 2D and 3D media. We present
simple test cases (which can be treated as benchmarks), to show the nature
of these solutions. In all the calculations we assume the atmosphere to
be isothermal.

We organize our discussions in terms of two effects. One is 
macroscopic in nature--namely the effect of RT on the Stokes profiles
formed in 2D and 3D media. 
Another is microscopic in nature--namely the effect of an 
oriented weak magnetic field on line scattering (Hanle effect). 
We discuss how these two effects act together on the polarized
line formation.

\subsection{The Stokes profiles formed due to resonance scattering
in 2D and 3D media}
\label{results1}
A discussion on the behavior of Stokes profiles formed in 1D
media with PRD scattering can be found in \citet{mf88} and \citet{knnetal99}.
In Paper II, the nature of profiles in a 2D semi-infinite 
medium is compared with those formed in 1D semi-infinite medium 
for CRD and PRD scattering (see Figures 8 and 9 of Paper II). 
Here we discuss the emergent, spatially averaged 
$\bm{\mathcal {I}}$ and $(I, Q/I, U/I)$ in 2D
and 3D media for PRD scattering.

Figures~\ref{fig-2d-ikq} and \ref{fig-3d-ikq} show the frequency
dependence of the components of emergent, spatially averaged 
$\bm{\mathcal{I}}$ in 2D and 3D media
respectively. The model parameters are, $T_X=T_Y=T_Z=T=2\times10^{9}$,
$a=10^{-3}$, $\Gamma_E/\Gamma_R=10^{-4}$, $\Gamma_I/\Gamma_R=10^{-4}$,
$\kappa_c/\kappa_l=10^{-7}$, and $\mu=0.11$. 
Our choice of collisional parameters represent a situation
in which $r_{\rm II}$ type scattering dominates.
Different curves in each panel represent different 
radiation azimuths $\varphi_i(i=1,12)$=$60$, $45$, 
$30^{\circ}$, $300^{\circ}$, $315^{\circ}$, $330^{\circ}$, 
$120^{\circ}$, $135^{\circ}$, $150^{\circ}$, $240^{\circ}$, 
$225^{\circ}$, $210^{\circ}$. 

$I^0_0$ is the largest of all the components. For the chosen model
parameters, all the other non-zero components are of the same order 
of magnitude.
The components $I^{2,\rm x}_1$ and $I^{2,\rm y}_2$ are zero in a 
2D geometry due to symmetry reasons (see Appendix B of Paper II 
for a proof).

The $\varphi$ dependence of the $\bm{\mathcal {I}}$ comes from the
$\varphi$ dependence of the scattering phase matrix ($\hat{\Psi}$) 
elements. The spatial distribution of $\bm{\mathcal {I}}$, on the top
surface depends sensitively on the
monochromatic optical depths for the ray at these spatial points.
This is a transfer effect within the medium, for the chosen ray
direction. In the line core frequencies ($x \le 3$), the monochromatic
optical depths are larger, resulting in a relatively uniform spatial
distribution of $\bm{\mathcal {I}}$ on the top surface. The
$\varphi$ dependence appears as either symmetric or anti-symmetric
with respect to the $X$-axis from which $\varphi$ is measured.
Thus the spatial averaging leads to a weak dependence of 
$\bm{\mathcal {I}}$ on the azimuth angle $\varphi$. When the averaging 
is performed over sign changing quantities like the polarization
components, it leads to cancellation, resulting in vanishing of these
components.

The $\varphi$ dependence of $\bm{\mathcal {I}}$ in the line wings
can be understood by considering the action of the first column elements
of the $\hat{\Psi}$ matrix on $I^0_0$, which is the largest
among all the components. The elements of $\hat{\Psi}$ matrix are
listed in Appendix~\ref{appendixd}. 
$I^0_0$ is independent of $\varphi$ because it is controlled
by the element $\Psi_{11}$ which takes a constant value unity. 
Similarly $I^2_0$ is controlled by $\Psi_{21}$ which is also independent
of $\varphi$. However we see a weak $\varphi$ 
dependence of $I^2_0$ in the wings, which is due to the 
coupling of the last 4 components to $I^2_0$, which are
of equal order of magnitude as $I^2_0$, and are sensitive to the values
of $\varphi$.
The $\varphi$ dependence of $I^{2,{\rm y}}_1$
and $I^{2,{\rm x}}_2$ elements in both 2D and 3D geometries
is controlled by $\sin \varphi$ and $\cos 2\varphi$ functions appearing
in $\Psi_{41}$ and $\Psi_{51}$ elements respectively. 
The distribution of angle points $\varphi$ in Carlson B quadrature
is such that among the 12 $\varphi$ values in the grid,
$\sin \varphi$ takes only 6 distinct values, and $\cos 2\varphi$ takes
only 3 distinct values (see Table~\ref{table_1}).
The components $I^{2,{\rm x}}_1$ and $I^{2,{\rm y}}_2$ are non-zero
in 3D geometry unlike the 2D case. Their magnitudes are 
comparable to those of $I^{2,{\rm y}}_1$
and $I^{2,{\rm x}}_2$. The $\varphi$ dependence of these components
are controlled by $\cos \varphi$ and $\sin 2\varphi$ functions appearing
in $\Psi_{31}$ and $\Psi_{61}$ elements. In the far wings, all the
components of $\bm{\mathcal {I}}$ go to their continuum values, as
shown in the inset panels of Figures~\ref{fig-2d-ikq} and \ref{fig-3d-ikq}.
In a 1D geometry $I_0^0$ reaches the value of $B_{\lambda}$ (parameterized
as 1 here) in the far wings where the source function is dominated by
$B_{\lambda}$. This is because of the fact that the formal solution with
$B_{\lambda}$ as source function along a given ray leads to terms of the form 
$B_{\lambda}[1-\exp{(-\tau_{x,\rm max})}]$. In 1D medium 
$\tau_{x,\rm max}=T \kappa_{\rm tot}/\mu$. This implies that for 
semi-infinite 1D medium, $\exp{(-\tau_{x,\rm max})}=0$ so that
$I^0_0=B_{\lambda}$ in the far wings. However in semi-infinite 
2D and 3D media the distances traveled by the rays in a given 
direction at different spatial points on the top surface are 
not always the same and therefore $\exp{(-\tau_{x,\rm max})}$ is 
not always zero unlike the 1D case. 
Further the radiation drops sharply near the edges due to
finiteness of the boundaries. Therefore when we perform spatial averaging 
of emergent $I_0^0$ over such different spatial points on
the top surface of a 2D medium (which is actually a line), 
$I_0^0$ will take a value smaller than $B_{\lambda}$. For 
a similar reason (averaging over a plane) the value of $I_0^0$ in
the far wings in a 3D medium becomes even smaller than the value 
in a 2D medium. All other components reach zero in the far wings because
the radiation is unpolarized in the far wings (because of an unpolarized 
continuum).

\begin{figure*}
\centering
\includegraphics[scale=0.9]{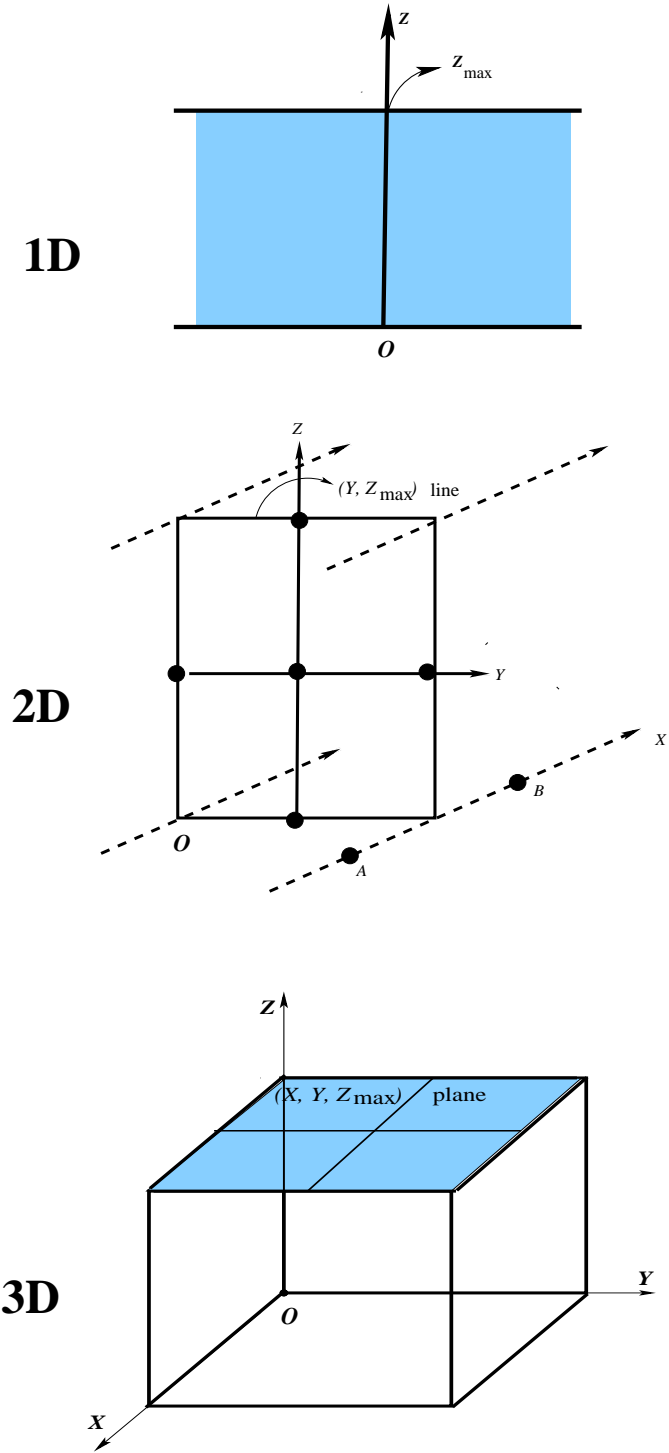}
\caption{The RT in 1D, 2D and 3D geometries. The $Z_{\rm max}$, 
$(Y,Z_{\rm max})$, and $(X,Y,Z_{\rm max})$ represent respectively, the point, 
the line, and the plane on which the emergent solutions are shown in 
this paper. The corresponding atmospheric reference frame is shown in 
Figure~\ref{fig-scatgeo}. The points A
and B marked on the 2D geometry figure represent an example of 
the spatial points where the symmetry of the polarized radiation field 
(Equation~\ref{2D-symmetry}) is valid in a 2D medium.}
\label{fig-geometry}
\end{figure*}

\begin{figure*}
\centering
\includegraphics[scale=0.5]{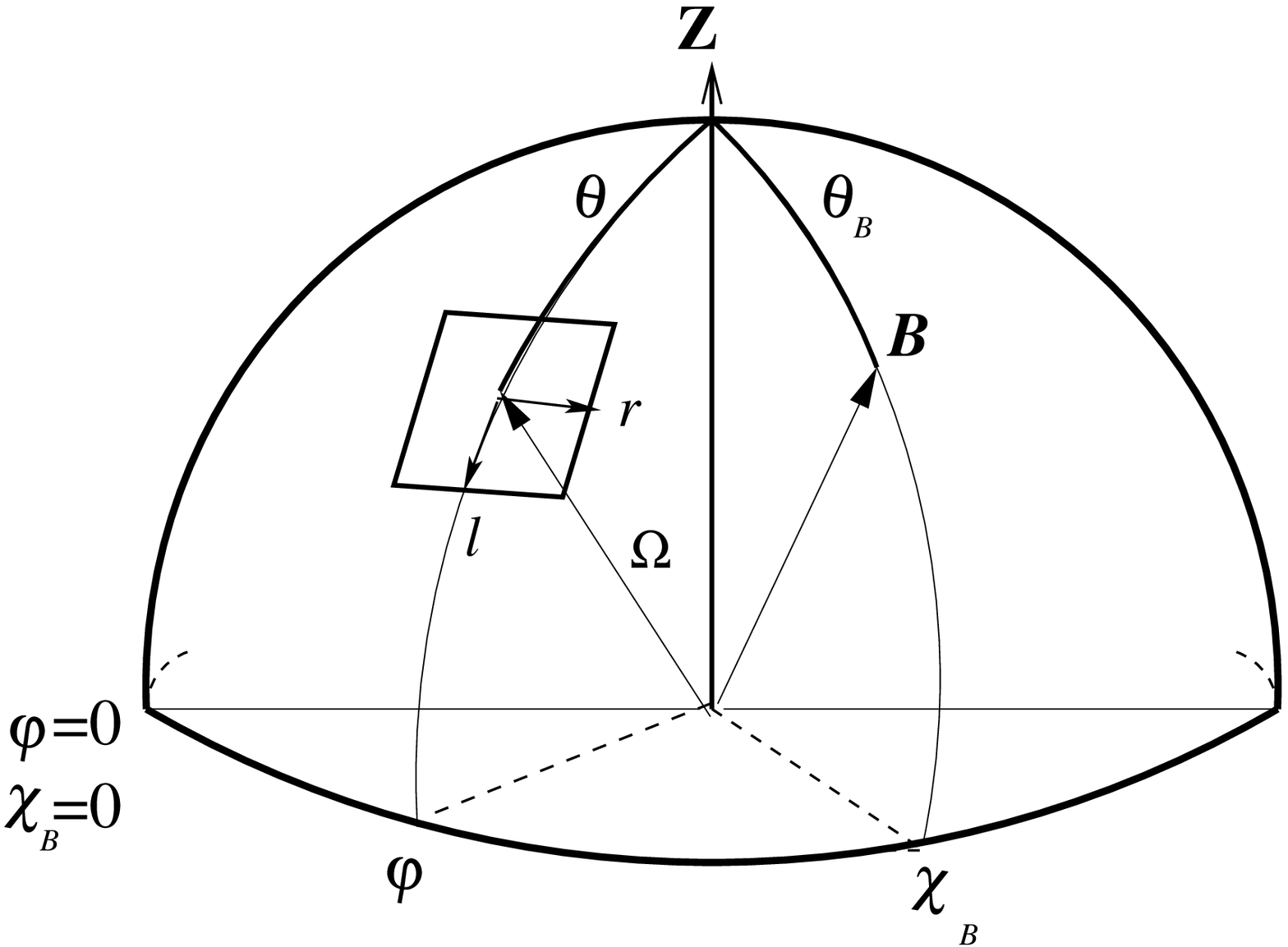}
\caption{The atmospheric reference frame.
The angle pair $(\theta,\varphi)$ define
the outgoing ray direction. The magnetic field is 
characterized by $\bm{B}=(\Gamma,\theta_B,\chi_B)$, where $\Gamma$ is 
the Hanle efficiency parameter and ($\theta_B,\chi_B$) 
defines the field direction. $\Theta$ is the scattering angle.}
\label{fig-scatgeo}
\end{figure*}

\begin{figure*}
\centering
\includegraphics[scale=0.5]{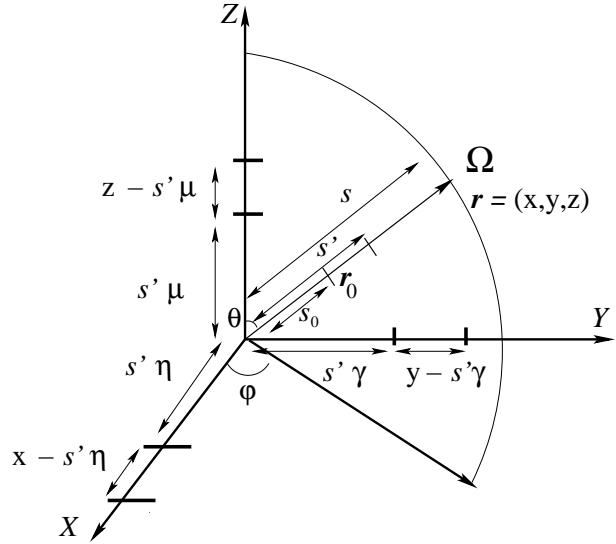}
\caption{The definition of the position vector $\bm{r}$ and the
projected distances $\bm{r}-s'\bm{\Omega}$ which appear in 
Equation~\ref{3d-formal}. $\bm{r}_0$ and $\bm{r}$ are the 
arbitrary initial and final locations that appear in formal 
solution integral (Equation~\ref{3d-formal}).}
\label{fig-fs}
\end{figure*}

\begin{figure*}
\centering
\includegraphics[scale=0.4]{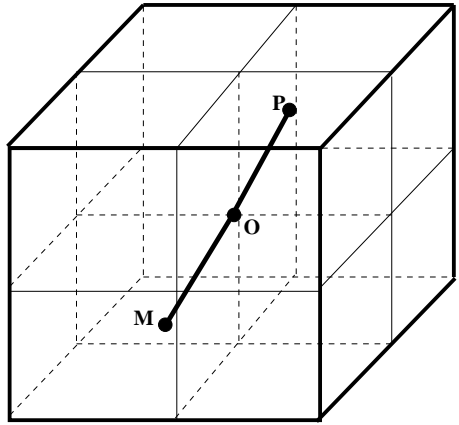}
\caption{An elemental cube, showing the transfer along
a section of the ray path, called a short characteristic 
(${\rm MOP}$). The quantities $\bm{\mathcal S}$, $\kappa_{\rm tot}$
at $\rm M$ and $\rm P$, and $\bm{\mathcal I}_M$ at $\rm M$ are
computed using parabolic interpolation formulae as $\rm M$ and $\rm P$
are non-grid points.}
\label{fig-sc}
\end{figure*}

\begin{figure*}
\centering
\includegraphics[scale=0.6]{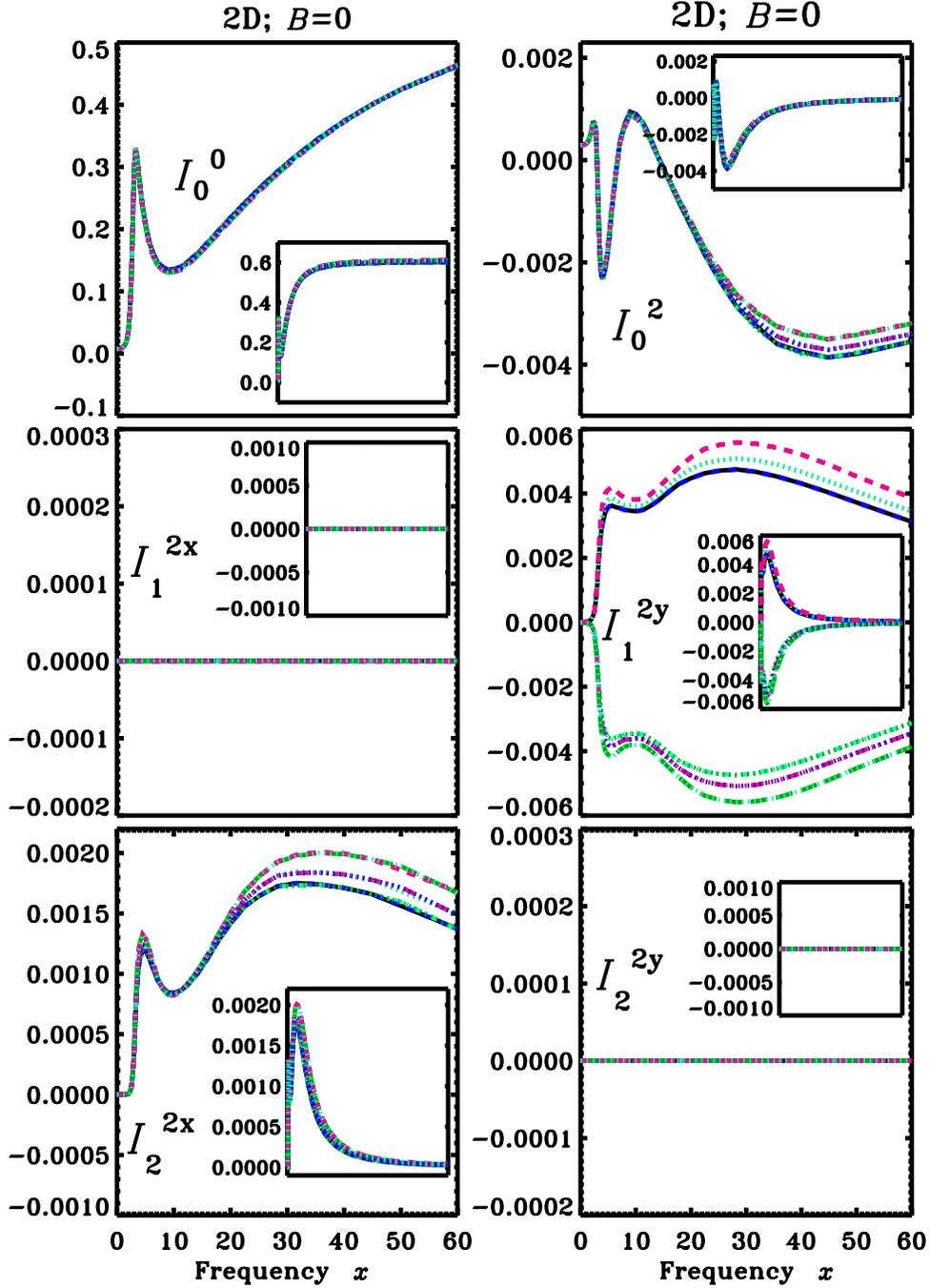}
\caption{The emergent, spatially averaged irreducible
Stokes vector components formed in a non-magnetic 2D medium.
Different curves represent different values of the radiation azimuth
$\varphi$. The value of $\mu=0.11$. The other model parameters are
given in Section~\ref{results1}. The inset panels show the far wing behavior
of $\bm{\mathcal {I}}$. The $x$ grid for these inset panels is
$0 \le x \le 600$.}
\label{fig-2d-ikq}
\end{figure*}

\begin{figure*}
\centering
\includegraphics[scale=0.6]{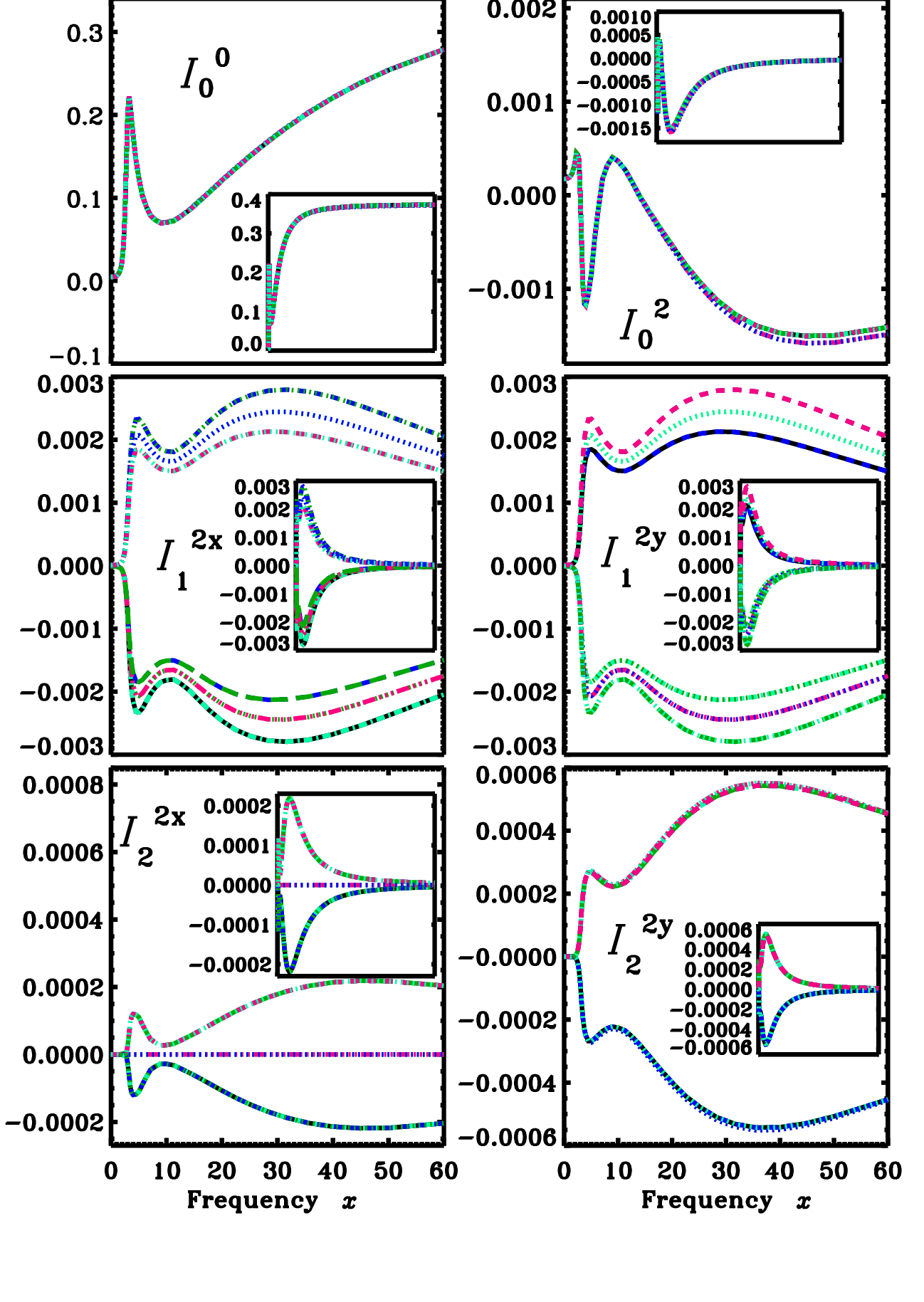}
\caption{Same as Figure~\ref{fig-2d-ikq} but for a 3D medium.}
\label{fig-3d-ikq}
\end{figure*}

The way in which the components of $\bm{\mathcal {I}}$ 
depend on $\varphi$ is different in
2D and 3D geometries (compare Figures~\ref{fig-2d-ikq} and \ref{fig-3d-ikq}).
This is a direct effect of spatial averaging. In a 2D medium, 
spatial averaging of the profiles is performed over the 
line $(Y,Z_{\rm max})$ marked in Figure~\ref{fig-geometry}, 
whereas in a 3D medium the averaging is performed
over the plane $(X,Y,Z_{\rm max})$ marked in Figure~\ref{fig-geometry}. The 
2D spatial averaging actually samples only a part of the 
plane considered for averaging in a 3D medium. 
Also, 2D geometry has an implicit assumption of front-back
symmetry of the polarized radiation field with respect to 
the infinite $X$ axis in the non-magnetic case, namely
\begin{eqnarray}
&&I(\bm{r},\theta,\varphi,x)=I(\bm{r},\theta,\pi-\varphi,x),\nonumber \\
&&I(\bm{r},\theta,\pi+\varphi,x)=I(\bm{r},\theta,2\pi-\varphi,x),
\nonumber \\
&&Q(\bm{r},\theta,\varphi,x)=Q(\bm{r},\theta,\pi-\varphi,x),\nonumber \\
&&Q(\bm{r},\theta,\pi+\varphi,x)
=Q(\bm{r},\theta,2\pi-\varphi,x),\nonumber \\
&&U(\bm{r},\theta,\varphi,x)=-U(\bm{r},\theta,\pi-\varphi,x),\nonumber \\
&&U(\bm{r},\theta,\pi+\varphi,x)=-U(\bm{r},\theta,2\pi-\varphi,x),
\nonumber \\
&&\theta \in [0,\pi], \varphi \in [0,\pi/2].
\label{2D-symmetry}
\end{eqnarray}
See Appendix B of Paper II for a proof of Equation~(\ref{2D-symmetry}).
However no such assumptions are involved in 3D geometry.

\begin{figure*}
\centering
\includegraphics[scale=0.45]{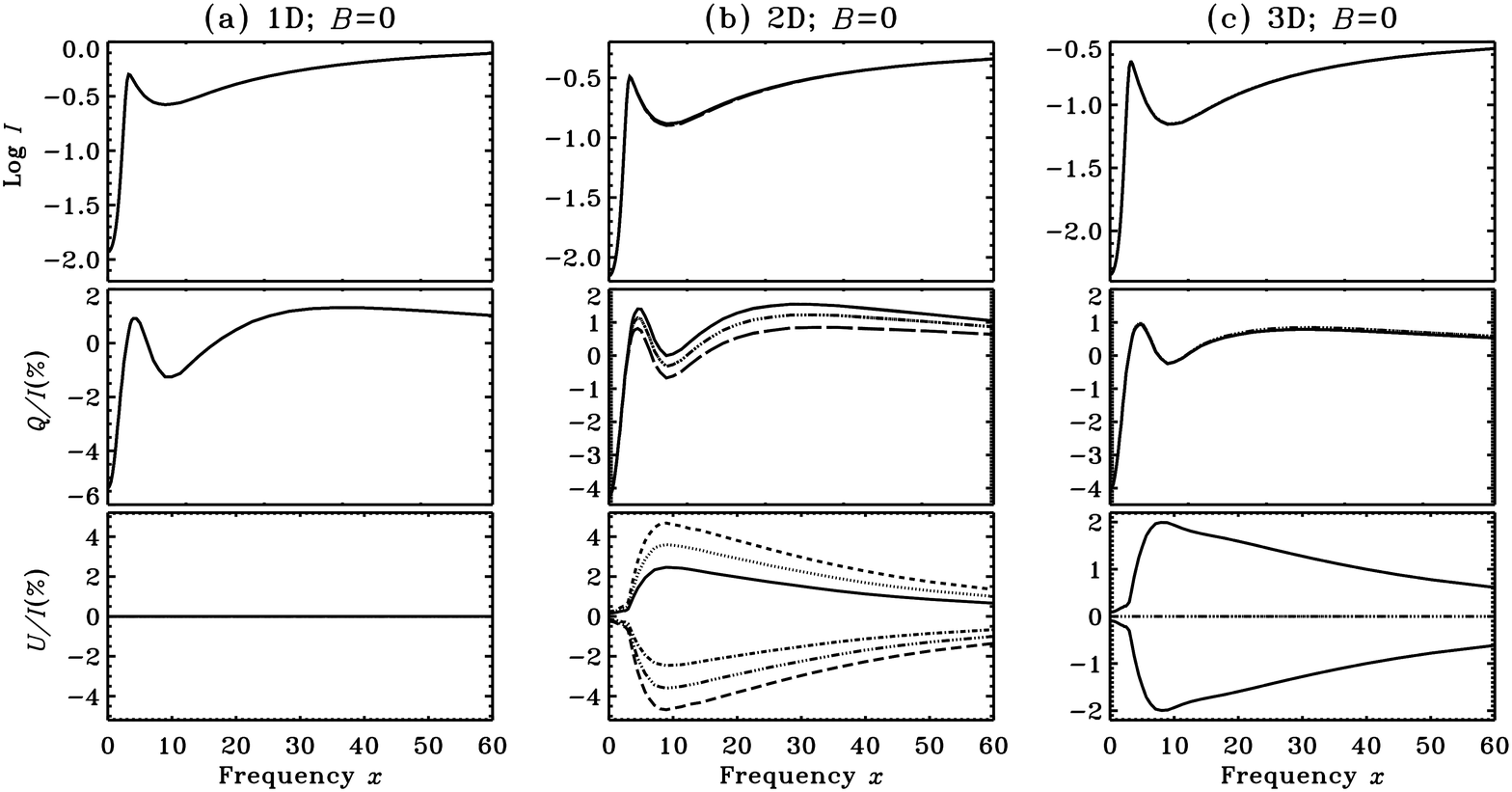}
\caption{The emergent, spatially averaged $(I, Q/I, U/I)$ 
in non-magnetic 1D, 2D and 3D media.
Different curves represent different values of the radiation azimuth
$\varphi$. The value of $\mu=0.11$. The other model parameters are
given in Section~\ref{results1}.}
\label{fig-2d-3d-iqu}
\end{figure*}

Figures~\ref{fig-2d-3d-iqu}(a), (b) and (c) show 
$I, Q/I, U/I$ profiles in non-magnetic 1D, 2D and 3D media.
Intensity $I$ decreases monotonically from 1D to the 3D case, 
because of the leaking of radiation through the finite boundaries in 
the lateral directions which is specific to RT in 2D and 3D geometries. 
In panels (b) and (c), 
different curves represent different $\varphi$ values. 
Only one curve is shown in panel (a), because of the axi-symmetry
of the radiation field in the 1D medium. 
For the same reason, $|U/I|_{\rm 1D}=0$. 
The $\varphi$ dependence of $|Q/I|_{\rm 2D, 3D}$ and 
$|U/I|_{\rm 2D. 3D}$ directly follow from those
of the components of $\bm{\mathcal {I}}$ shown in 
Figures~\ref{fig-2d-ikq} and \ref{fig-3d-ikq},
and their combinations (see Appendix~\ref{appendixa} in 
this paper where we list the formulae used to construct the Stokes vector 
($I$, $Q$, $U$)$^T$ from the irreducible components of $\bm{\mathcal{I}}$).
At the line center, $[U/I]_{\rm 2D, 3D}\sim 0$. This is because
$U/I$ is zero in large parts of the top surface and
the positive and negative values of $U/I$ at $x=0$ are nearly 
equally distributed in a narrow region near the edges. 
A spatial averaging of such a distribution 
leads to cancellation giving a net value of
$U/I$ approaching zero. This is not the case in wing
frequencies of the $U/I$ profile (see discussions 
in Section~\ref{results3} for spatial distribution of 
$Q/I$ and $U/I$).

\subsection{The Stokes profiles in 2D and 3D media in the presence of 
a magnetic field}
\label{results2}
\begin{figure*}
\centering
\includegraphics[scale=0.6]{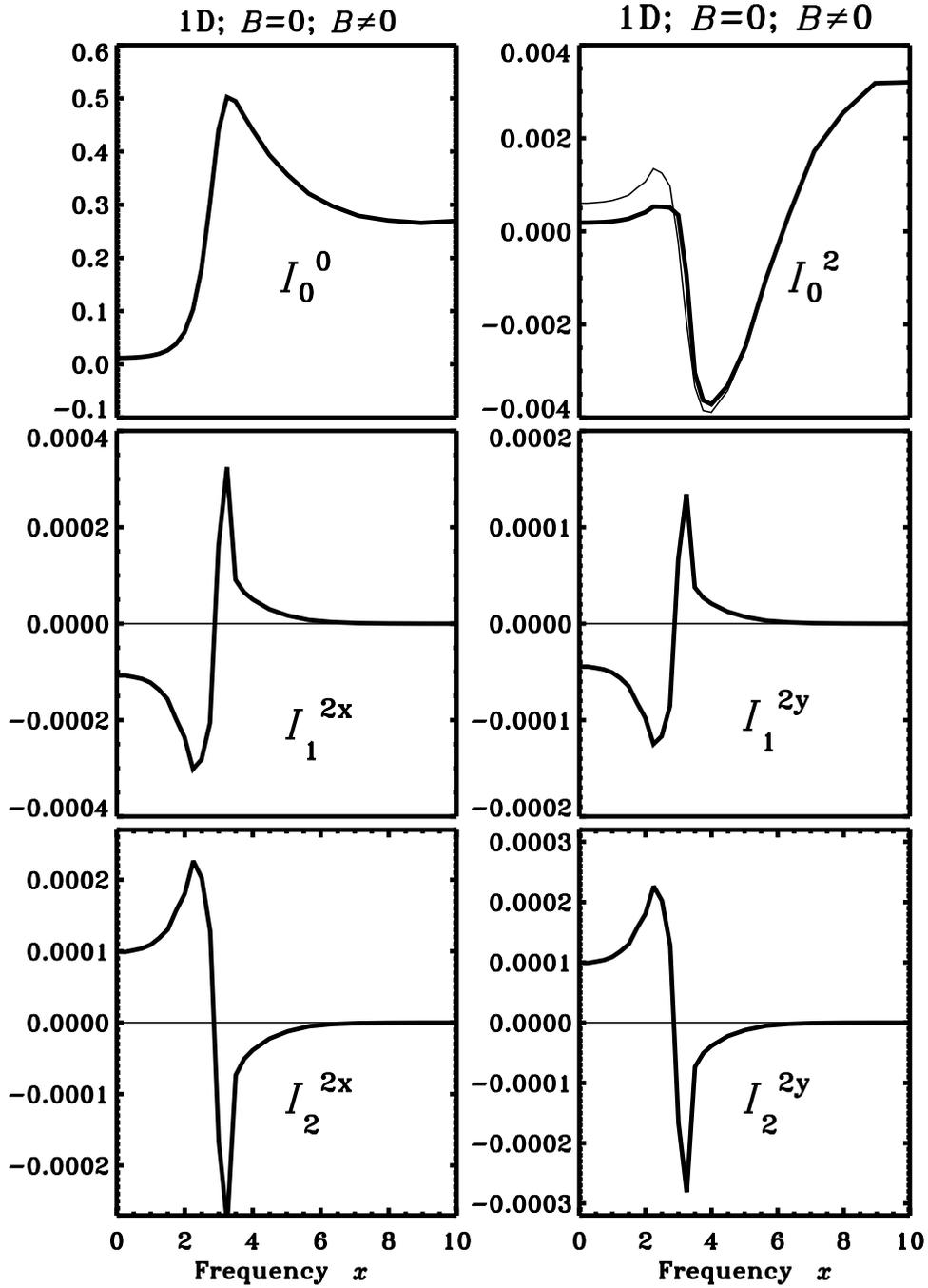}
\caption{Same as Figure~\ref{fig-2d-ikq} but for a magnetic
1D medium. The vector magnetic field
is represented by $(\Gamma,\theta_B,\chi_B)=(1,90^{\circ},68^{\circ})$. 
The thin solid lines show the corresponding non-magnetic
components.}
\label{fig-1d-mag-ikq}
\end{figure*}

\begin{figure*}
\centering
\includegraphics[scale=0.6]{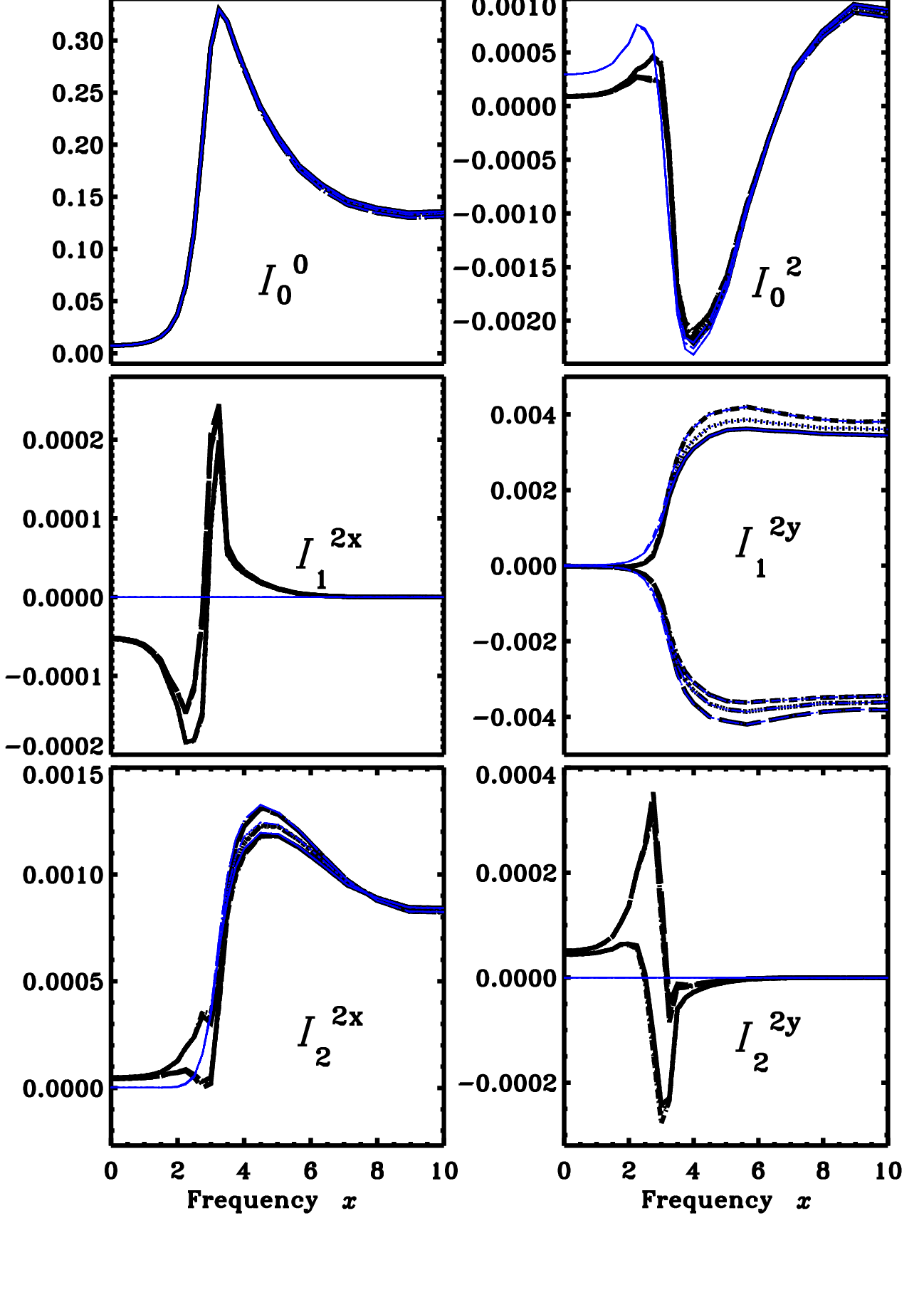}
\caption{Same as Figure~\ref{fig-1d-mag-ikq} but for a 2D
medium.}
\label{fig-2d-mag-ikq}
\end{figure*}

\begin{figure*}
\centering
\includegraphics[scale=0.6]{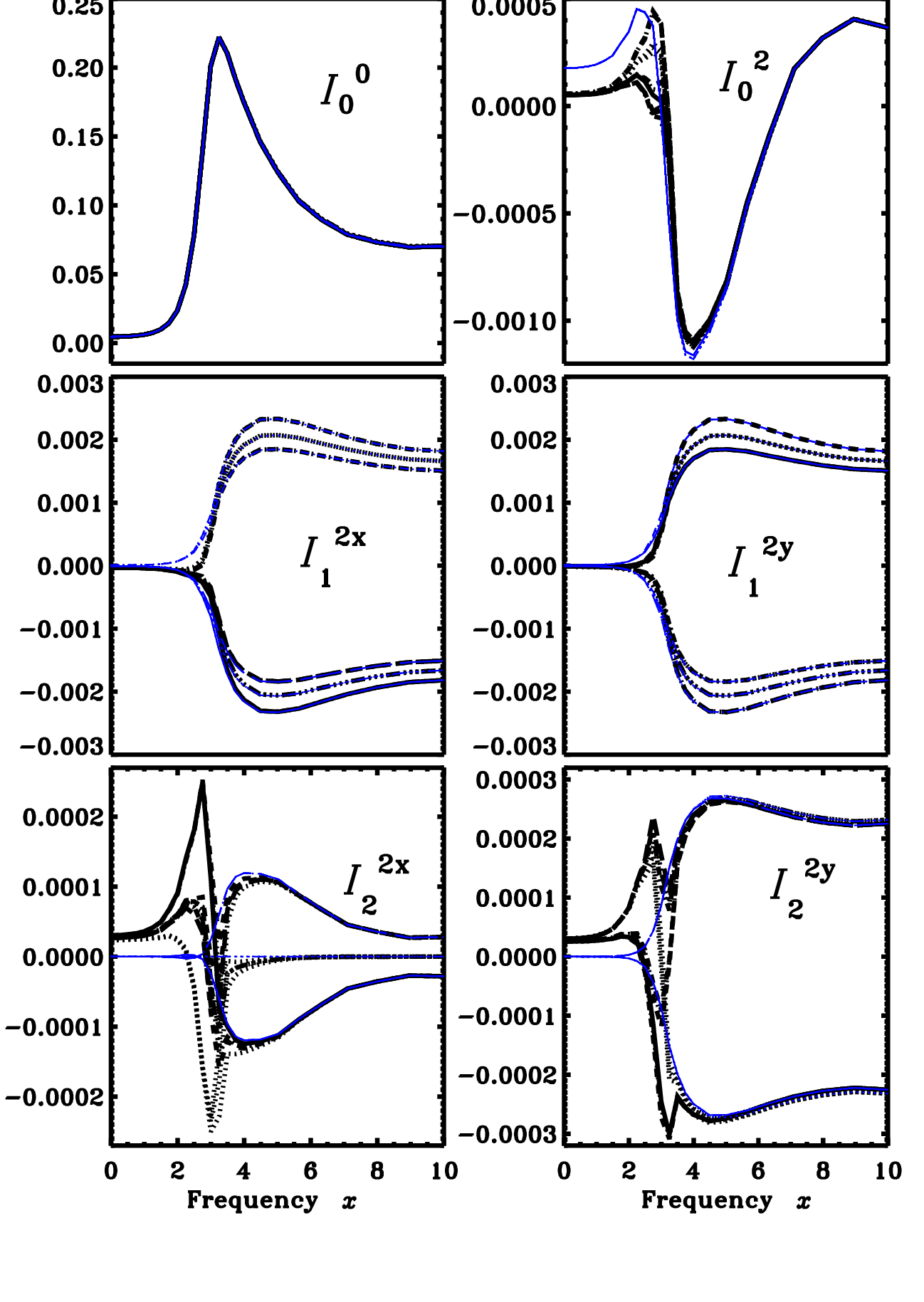}
\caption{Same as Figure~\ref{fig-1d-mag-ikq} but for a 3D
medium.}
\label{fig-3d-mag-ikq}
\end{figure*}

Figures~\ref{fig-1d-mag-ikq}, \ref{fig-2d-mag-ikq} and
\ref{fig-3d-mag-ikq} show all the 6 components of $\bm{\mathcal {I}}$
in magnetized 1D, 2D and 3D media respectively. The vector magnetic
field $\bm{B}$ is represented by 
$(\Gamma,\theta_B,\chi_B)=(1,90^{\circ},68^{\circ})$. 
The corresponding non-magnetic components are shown as thin solid
lines. Different line types in Figures~\ref{fig-2d-mag-ikq} 
and \ref{fig-3d-mag-ikq} correspond to different $\varphi$.
The irreducible components in 1D geometry 
are cylindrically symmetrical, even
when there is an oriented magnetic field.
Therefore there is only one curve in each panel in 
Figure~\ref{fig-1d-mag-ikq}.
When $\bm{B}=0$ the 4 components $I^{2{\rm x,y}}_{1,2}$ become 
zero due to axi-symmetry in 1D geometry (Figure~\ref{fig-1d-mag-ikq}). 
These components take non-zero values in the line core when 
$\bm{B}\neq0$. The magnitudes of $I^0_0$ and 
$I^2_0$ monotonically decrease
from 1D to 3D. In the 2D case, the 2 components which were 
zero when $\bm{B}=0$, take non-zero values in the line core, 
when $\bm{B}\neq0$. Unlike 1D geometry in 2D and 3D geometries,
a non-zero $\bm{B}$ causes the last 4 components to become 
sensitive to $\varphi$. The components $I^{2{\rm y}}_{1}$ 
in 2D and $I^{2{\rm x}}_{1}$ and $I^{2{\rm y}}_{1}$ in 3D 
remain almost unaffected by $\bm{B}$.
This behavior is particular to the present choice of $\bm{B}$. For a
different choice of $\bm{B}$, the behavior of the 6 components may 
differ from what is shown in these figures. In all the geometries, 
the components go to their non-magnetic (Rayleigh scattering) 
values in the wings, because the Hanle effect operates only in 
the line core region.

\begin{figure*}
\centering
\includegraphics[scale=0.53]{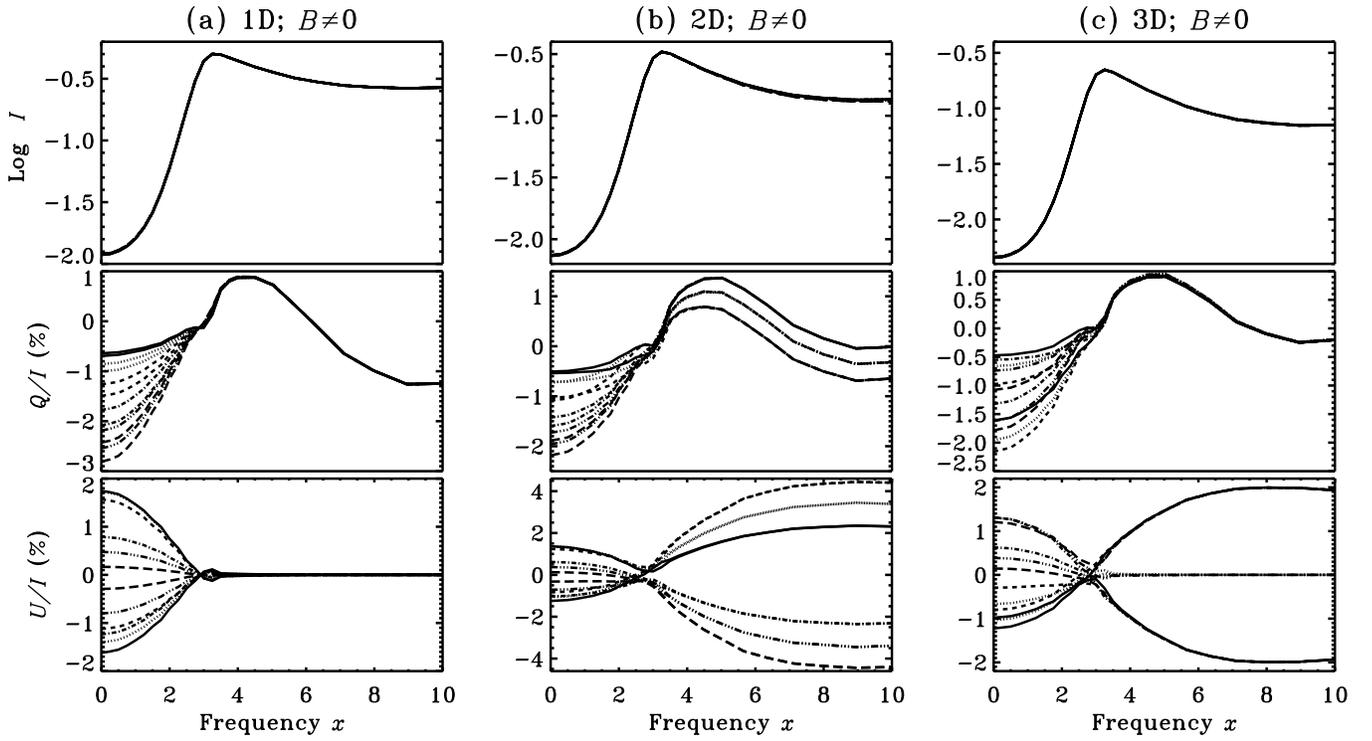}
\caption{A comparison of emergent $I$, $Q/I$ and $U/I$ profiles formed
in a magnetized 1D media with the emergent, 
spatially averaged $I$, $Q/I$ and $U/I$ 
formed in a magnetized 2D and 3D media. The
model parameters are same as in Figure~\ref{fig-1d-mag-ikq}.} 
\label{fig-1d-2d-3d}
\end{figure*}

Figures~\ref{fig-1d-2d-3d}(a), (b) and (c) show spatially
averaged $I$, $Q/I$, $U/I$ in 1D, 2D and 3D geometries respectively.  
Due to the finiteness of the boundaries in 2D and 3D media the value 
of spatially averaged $I$ decreases monotonically from 1D to 3D.
The dependence of $Q/I$ and $U/I$ on $\varphi$ 
in 1D medium is purely due to the $\varphi$ dependence
coming from the formulae used to convert $\bm{\mathcal {I}}$
to $I$, $Q$ and $U$ (see Appendix~\ref{appendixa}). 
In 2D and 3D media, the $\varphi$ dependence comes from both, the
$\varphi$ dependence of the respective components of $\bm{\mathcal {I}}$,
and also the above mentioned conversion formulae. 
The magnitudes of $Q/I$ and $U/I$
decrease in 2D and 3D geometries due to the spatial averaging process.
The wings of $Q/I$ and $U/I$ in 1D are insensitive to $\varphi$ due
to the inherent axi-symmetry. In 2D they become more sensitive
to $\varphi$ values. Again they become weakly sensitive to $\varphi$
in 3D geometry. These differences in sensitivities of $Q/I$, $U/I$
to the azimuth angle $\varphi$ in 2D and 3D geometries is due to
the way in which the spatial averaging is performed in these
geometries (see discussions above Equation~(\ref{2D-symmetry})).
\subsubsection{Polarization diagrams in 1D and 2D media}
\begin{figure*}
\centering
\includegraphics[scale=0.8]{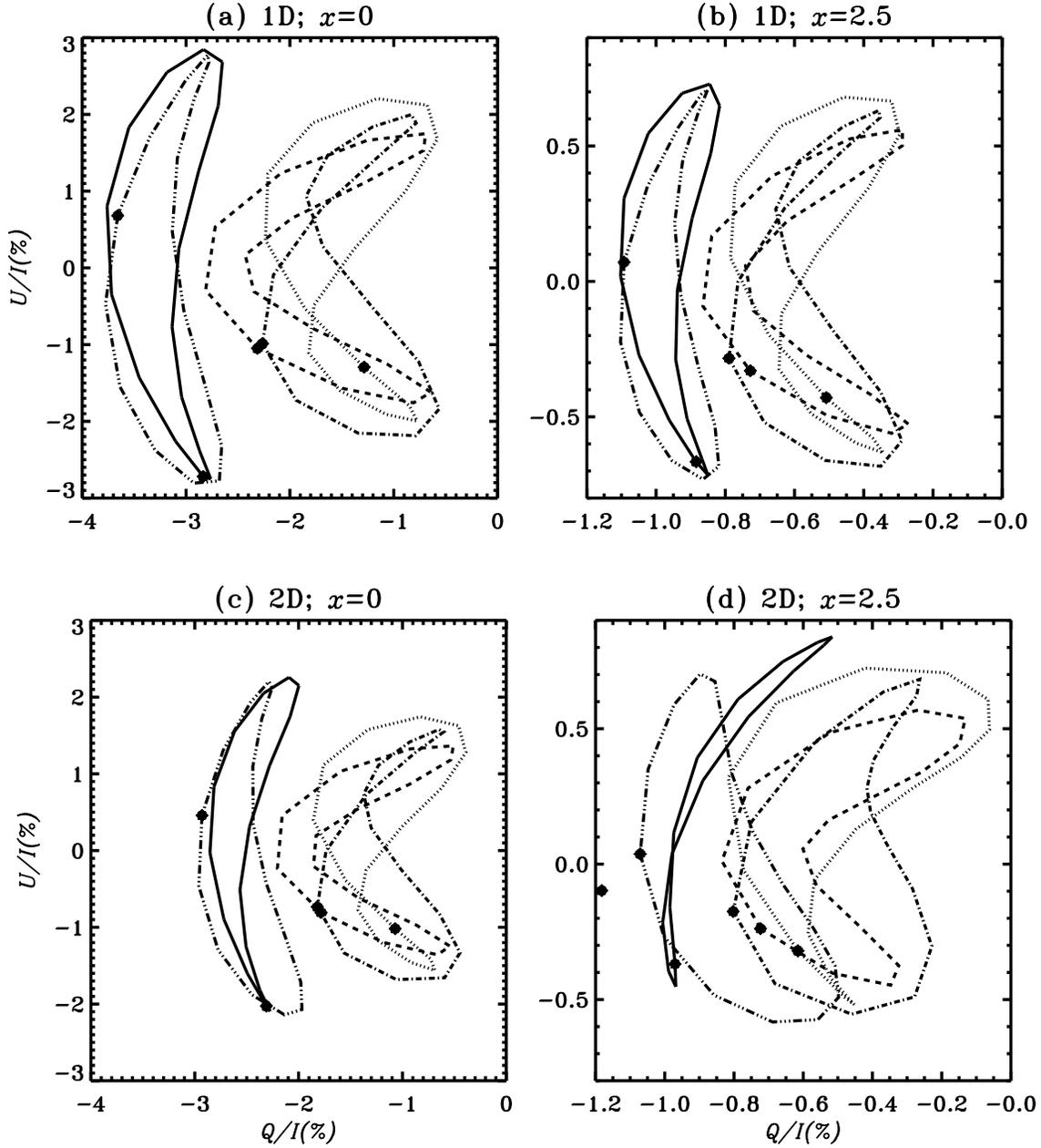}
\caption{A comparison of the polarization diagrams in 1D and 2D
media for two different values of frequency $x$. In 2D, the 
spatially averaged quantities are shown. The magnetic field parameters
are given by $\Gamma=1$, five values of $\theta_B$ in the range
$30^{\circ}$ to $150^{\circ}$ in steps of $30^{\circ}$, seventeen 
values of $\chi_B$ in the range $0^{\circ}$ to $360^{\circ}$ in 
steps of $22^{\circ}.5$. Different line types correspond to different
values of $\theta_B$. Heavy square symbol represents $\chi_B=0$, and
as we move in the counter-clockwise direction, $\chi_B$ takes
increasingly larger values. The ray direction is specified 
by $(\mu,\varphi)$=$(0.11,60^{\circ})$. The line types represent
different $\theta_B$, namely (solid, dotted, dashed, dot-dashed, 
dash-triple-dotted)=($30^{\circ}, 60^{\circ}, 90^{\circ}, 
120^{\circ}, 150^{\circ}$).}
\label{fig-poldiag}
\end{figure*}

In Figure~\ref{fig-poldiag} we show polarization diagrams
\citep[see e.g.,][]{jos94}, which are plots of
$Q/I$ versus $U/I$ for a given value of frequency $x$, ray
direction $(\mu,\varphi)$, and varying the field parameters two 
out of three at a time. We take $\Gamma=1$, and 
vary $\theta_B$ and $\chi_B$ values.
For the 2D case we show spatially averaged quantities. 

For $x=0$, the shapes of closed curves (loops) in the 
polarization diagrams are the same in both 1D and 2D cases. 
When compared to the loops in 1D, the sizes of the loops in 
2D are smaller by about 1\% in the magnitudes of $Q/I$ and $U/I$, 
which is due to spatial averaging. 

For $x=2.5$, the shapes of the the loops in 2D  
are quite different from those for 1D.
For e.g., the solid curve in panel (d) is narrower than
the one in panel (b) which correspond to $\theta_B=30^{\circ}$.
On the other hand, the dash-triple-dotted curve in panel (d)
is broader than the one in panel (b),
which correspond to $\theta_B=120^{\circ}$. 
The orientation of a given loop with respect to 
the vertical line $(Q/I=0)$ is a measure of the 
sensitivity of $(Q/I, U/I)$ to the field orientation
$\theta_B$. The size of a loop is a measure of the sensitivity
of $(Q/I, U/I)$ to the field azimuth $\chi_B$. 
The values of $|Q/I|_{2D}$ and $|U/I|_{2D}$ can be larger or smaller
than $|Q/I|_{1D}$ and $|U/I|_{1D}$ for $x=2.5$.
The sensitivity of the line wing ($x=2.5$) polarization 
to $(\theta_B,\chi_B)$ is different in 1D and 2D geometries,
when compared to the sensitivity of line center ($x=0$) 
polarization. This is because at $x=0$ we sample
mainly the outermost layers of the semi-infinite media.
At $x=2.5$ we actually sample internal inhomogeneities
of the radiation field in $(Y,Z)$ directions
in the 2D case, and only those in the $Z$ direction, in the
1D case. We have noticed that the spatial distribution
of $Q/I, U/I$ at $x=0$ is relatively more homogeneous,
than at $x=2.5$ (see figures and discussions
in Section~\ref{results3} for spatial distribution of
$Q/I$ and $U/I$).
\subsection{The spatial variation of emergent $(Q/I, U/I)$ in a 3D medium}
\label{results3}
\begin{figure*}
\centering
\includegraphics[scale=0.25]{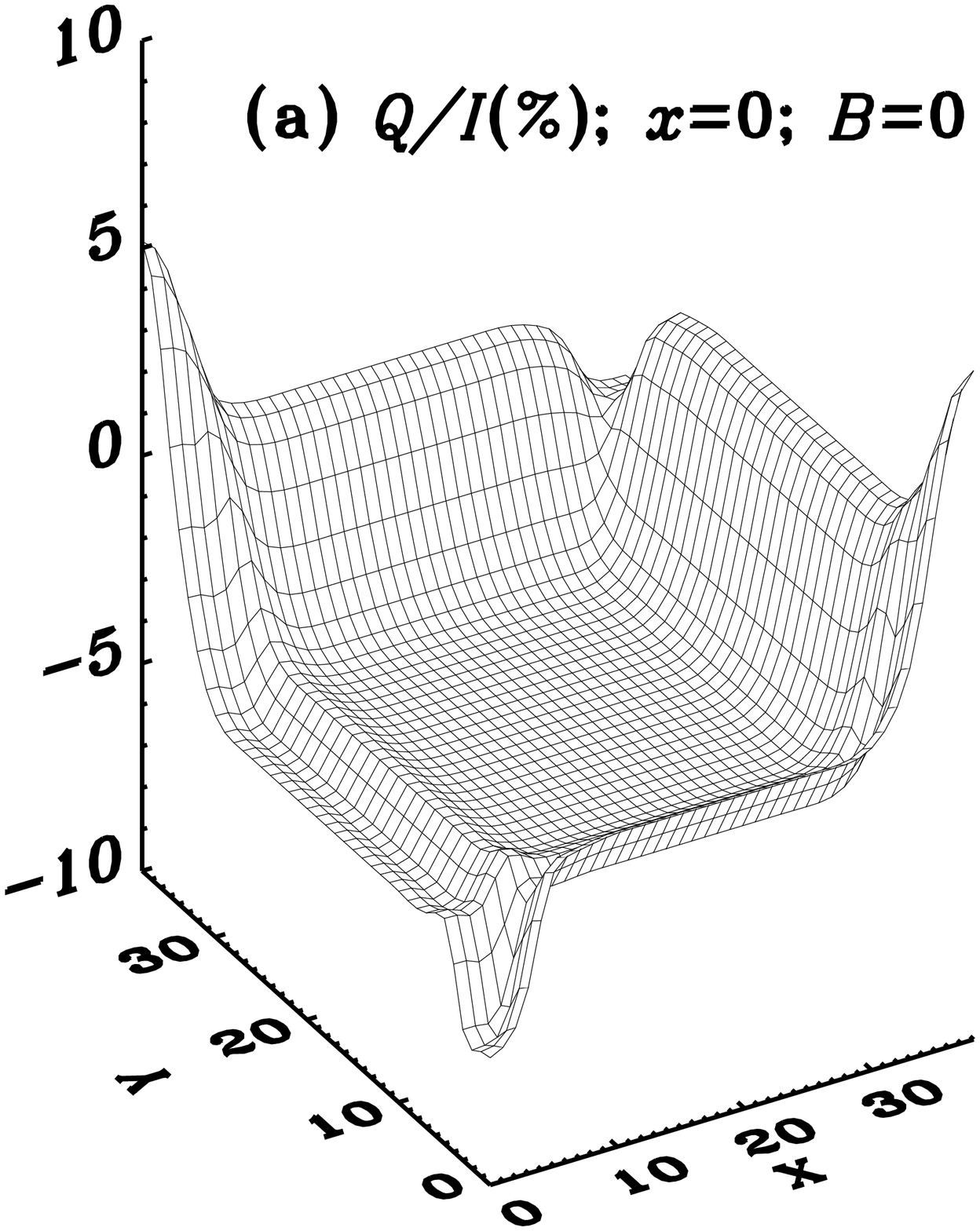}
\includegraphics[scale=0.25]{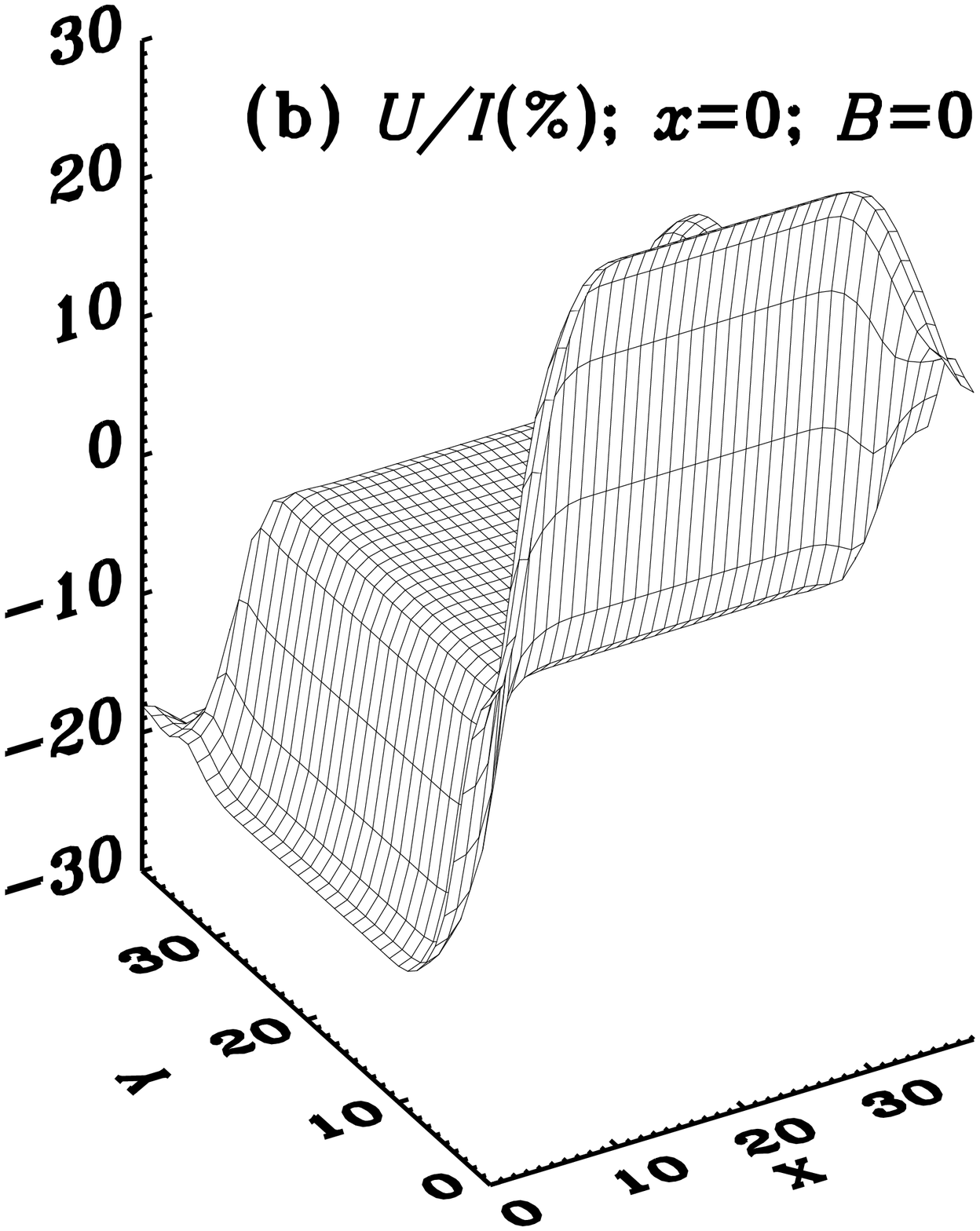}
\includegraphics[scale=0.25]{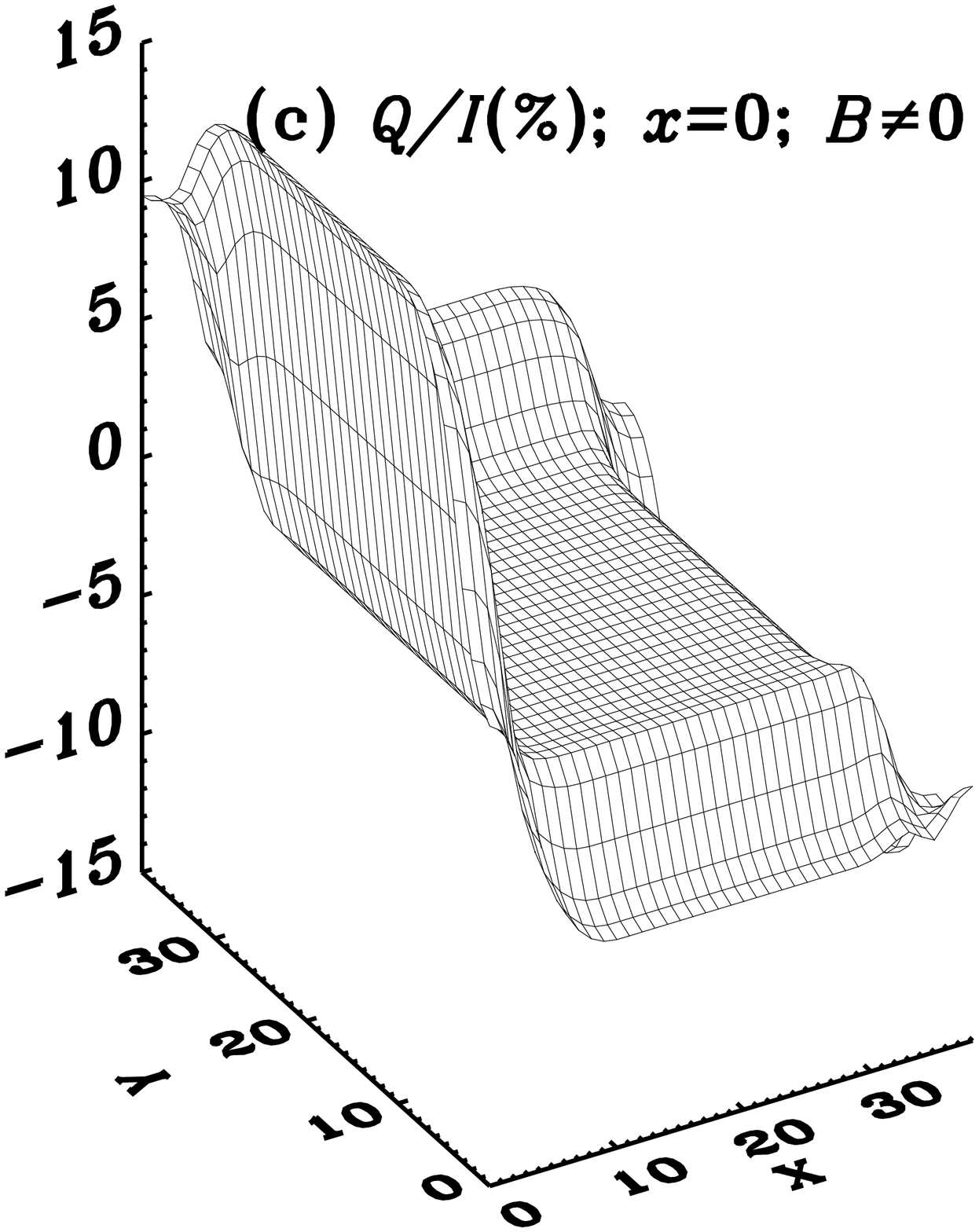}
\includegraphics[scale=0.25]{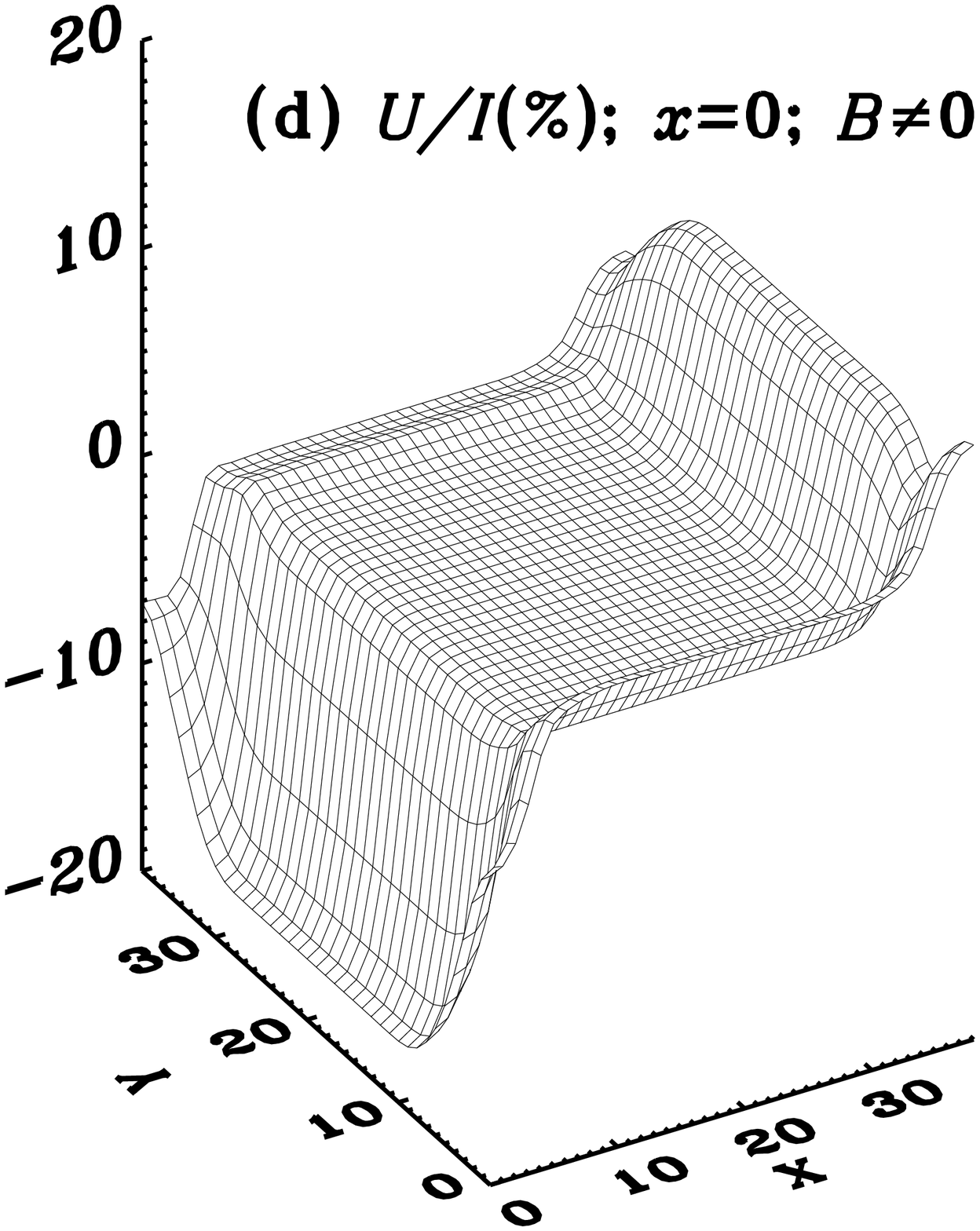}
\includegraphics[scale=0.25]{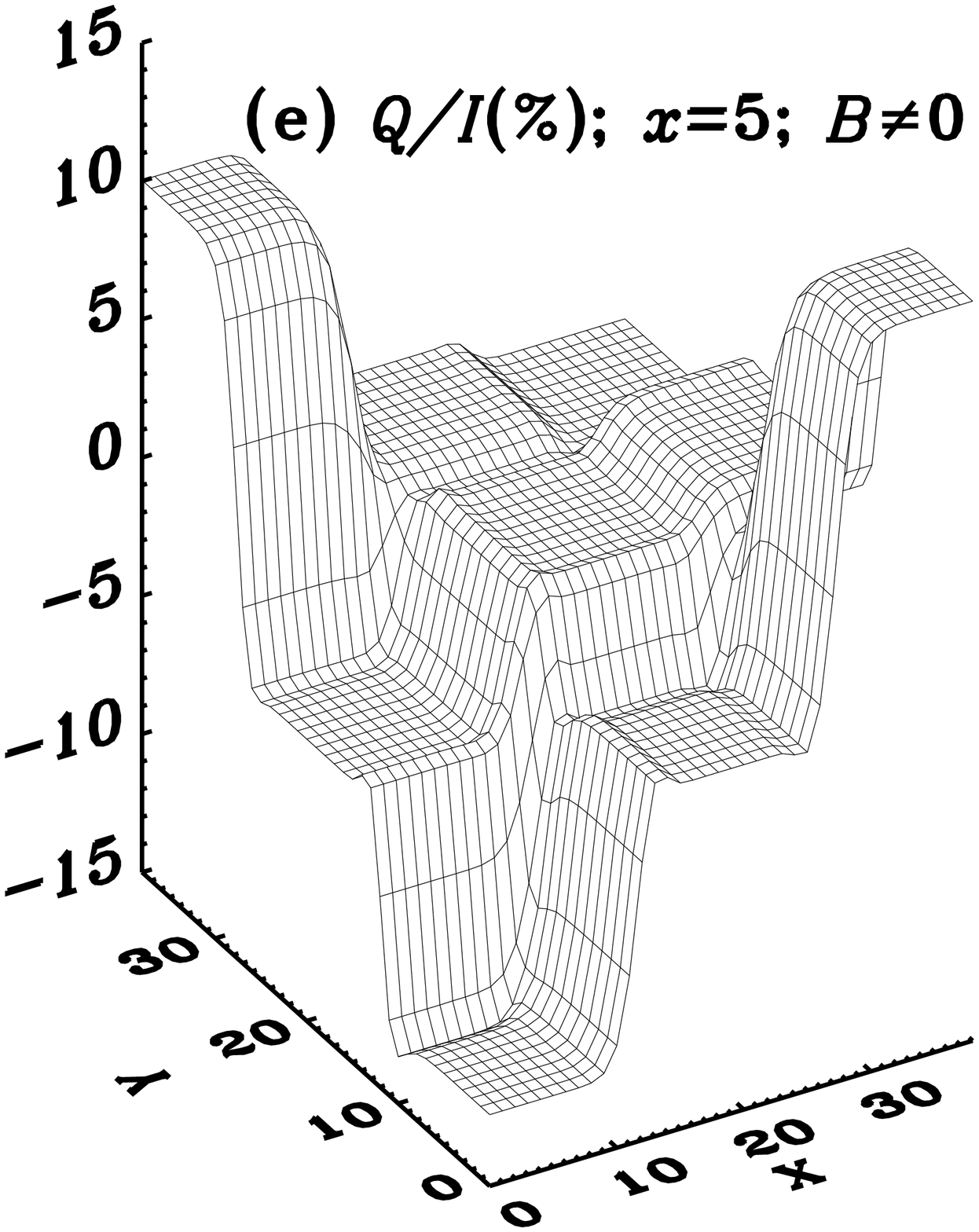}
\includegraphics[scale=0.25]{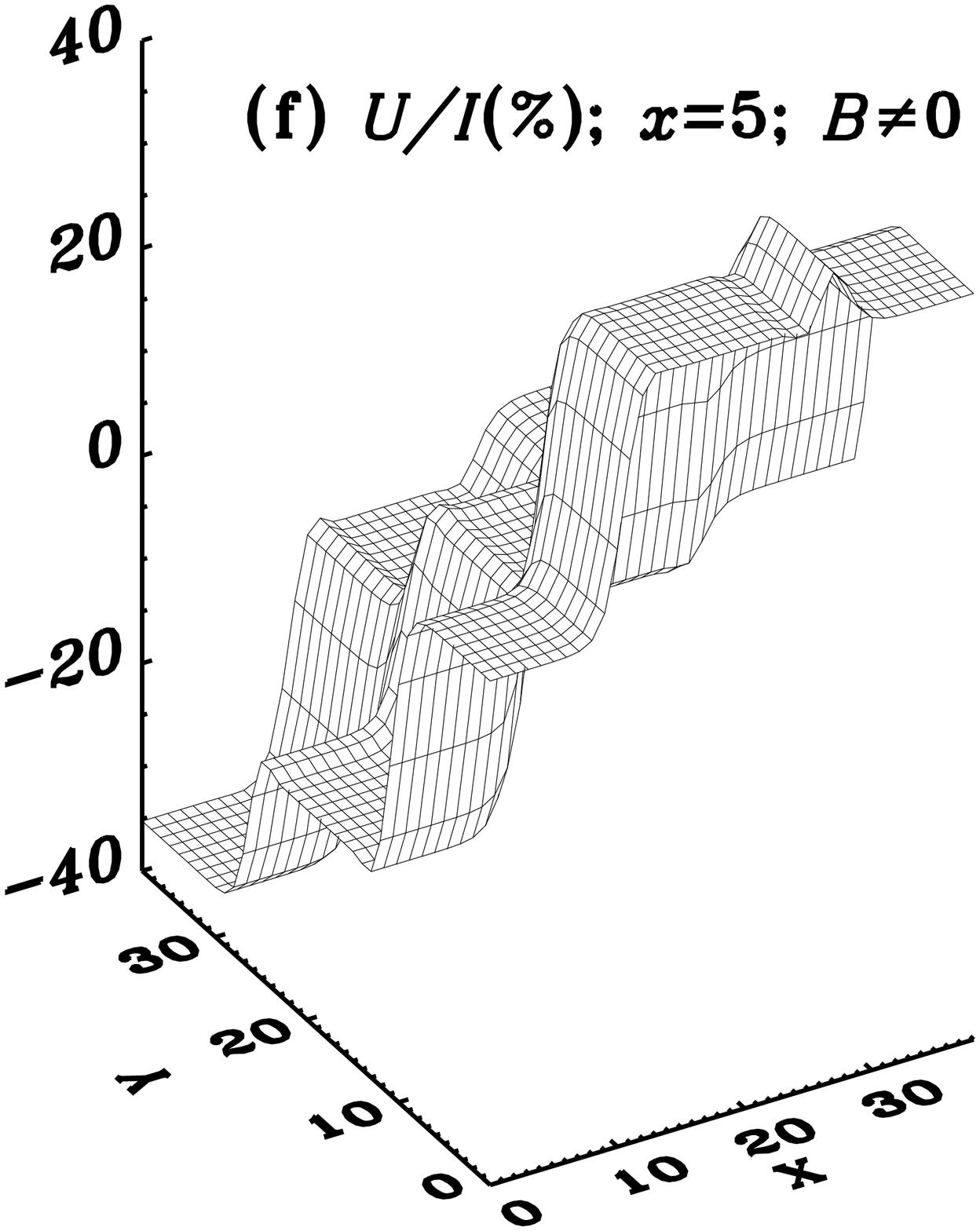}
\caption{The spatial distribution of ($Q/I, U/I$) on the
top surface of a 3D medium. The $Q/I$ and $U/I$ are 
plotted as a function of the grid indices of $\tau_X$
and $\tau_Y$. 
The ray direction is specified
by $(\mu,\varphi)$=$(0.11,60^{\circ})$.
Panels (a) and (b) demonstrate
purely the multi-D effects. Panels (c) and (d) demonstrate the magnetic field
effects. Panels (e) and (f) demonstrate the PRD effects. 
See Section~\ref{results3} for details.}
\label{fig-surface}
\end{figure*}

In Figure \ref{fig-surface} we show surface plots of $Q/I$
and $U/I$ formed in a 3D media. The region chosen for showing the 
spatial distribution is the top surface plane ($X,Y,Z_{\rm max}$).

Figures~\ref{fig-surface}(a), (b)
demonstrate purely the effects of multi-D geometry on the 
$(Q/I, U/I)$ profiles.
In Figure~\ref{fig-surface}(a) $Q/I$ shows a homogeneous distribution
at the interiors of the top surface (away from the boundaries) 
approaching a constant value ($\sim -3.6\%$). 
Large parts of the top surface contribute to the negative 
values of $Q/I$ and only a narrow region
near the edges contribute to positive values.
The magnitudes of $Q/I$ sharply raise
near the edges. This is due to the finite boundaries
of the 3D medium. Maximum value of $|Q/I|$ in these figures
is $\sim 6\%$. In Figure~\ref{fig-surface}(b) $U/I$ is nearly zero 
at the interiors of the top surface.
Near the edges, the values of $U/I$ sharply raise
and $|U/I|$ takes a maximum value of $\sim 20\%$.

Figures~\ref{fig-surface}(c), (d)
demonstrate the effects of magnetic field on the $(Q/I, U/I)$ profiles.
The magnetic field vector is represented by 
$\bm{B}$=$(\Gamma, \theta_B, \chi_B)$=$(1, 30^{\circ}, 68^{\circ})$.
The nature of homogeneity at the interior and sharp raise
near the edges of the 3D surface, in the values of $Q/I$ and $U/I$
remain similar in both the magnetic and non-magnetic cases.
An important effect of $\bm{B}$ is to significantly change the 
values of $Q/I$ and $U/I$ with respect to their
non-magnetic values. $|Q/I|$ values are slightly 
reduced at the interior and $Q/I$ now becomes $-2.3\%$.
Near the edges $|Q/I|$ is significantly enhanced and takes 
a maximum value of 15\%.
The interior values of $|U/I|$ continue to be nearly zero. 
The $|U/I|$ is reduced at different rates near different edges.
Now the maximum value of $|U/I|$ is 17\%.
We note that in 1D geometry, for $\mu=0.11$, 
any magnetic field configuration always causes a 
decrease in $|Q/I|$ and a fresh generation of $|U/I|$
with respect to the non-magnetic values.

Figures~\ref{fig-surface}(e), (f)
demonstrate the effects of PRD on the $(Q/I, U/I)$ profiles.
For this purpose we have chosen a wing frequency $x=5$.
The spatial distribution of $Q/I$ and $U/I$ is highly
inhomogeneous at the wing frequencies. 
This effect can be easily seen by comparing 
Figure~\ref{fig-surface}(a) which exhibits large spatial homogeneity
for $x=0$, with Figure~\ref{fig-surface}(e) 
which exhibits large spatial inhomogeneity
for $x=5$. For $x=0$, the optical depth of the medium
is large and therefore the radiation field in the line core
becomes homogeneous over large volumes of the cube.
The spatial inhomogeneity of the $Q/I$ at $x=5$ is actually caused by
the the nature of PRD function used in our
computations (which is dominated by the $r_{\rm II}$ function).
Due to the frequency coherent nature of $r_{\rm II}$, the photons
scattered in the wings get decoupled from the line core radiation
field. As the optical depth of the medium in the line wings
is smaller than in the line core, the wing radiation field becomes more
inhomogeneous and more polarized. Same arguments
are valid for the inhomogeneous distribution of $U/I$ on the top
surface of the 3D cube. This can be seen by comparing 
Figure~\ref{fig-surface}(b) 
with Figure~\ref{fig-surface}(f).
We recall that under the assumption of CRD, 
the values of $Q/I$ and $U/I$ are zero in the line wings
(see Figure 9 of Paper II for a comparison of emergent, spatially
averaged $Q/I$, $U/I$ profiles for CRD and PRD in a multi-D medium). 
The sharp increase in magnitudes of $Q/I$ 
and $U/I$ near the edges is larger for $x=5$ 
when compared to those for $x=0$.
Maximum value of $|Q/I|$  is now 10\% and that of $|U/I|$ is 40\%.

\begin{figure*}
\centering
\includegraphics[scale=0.25]{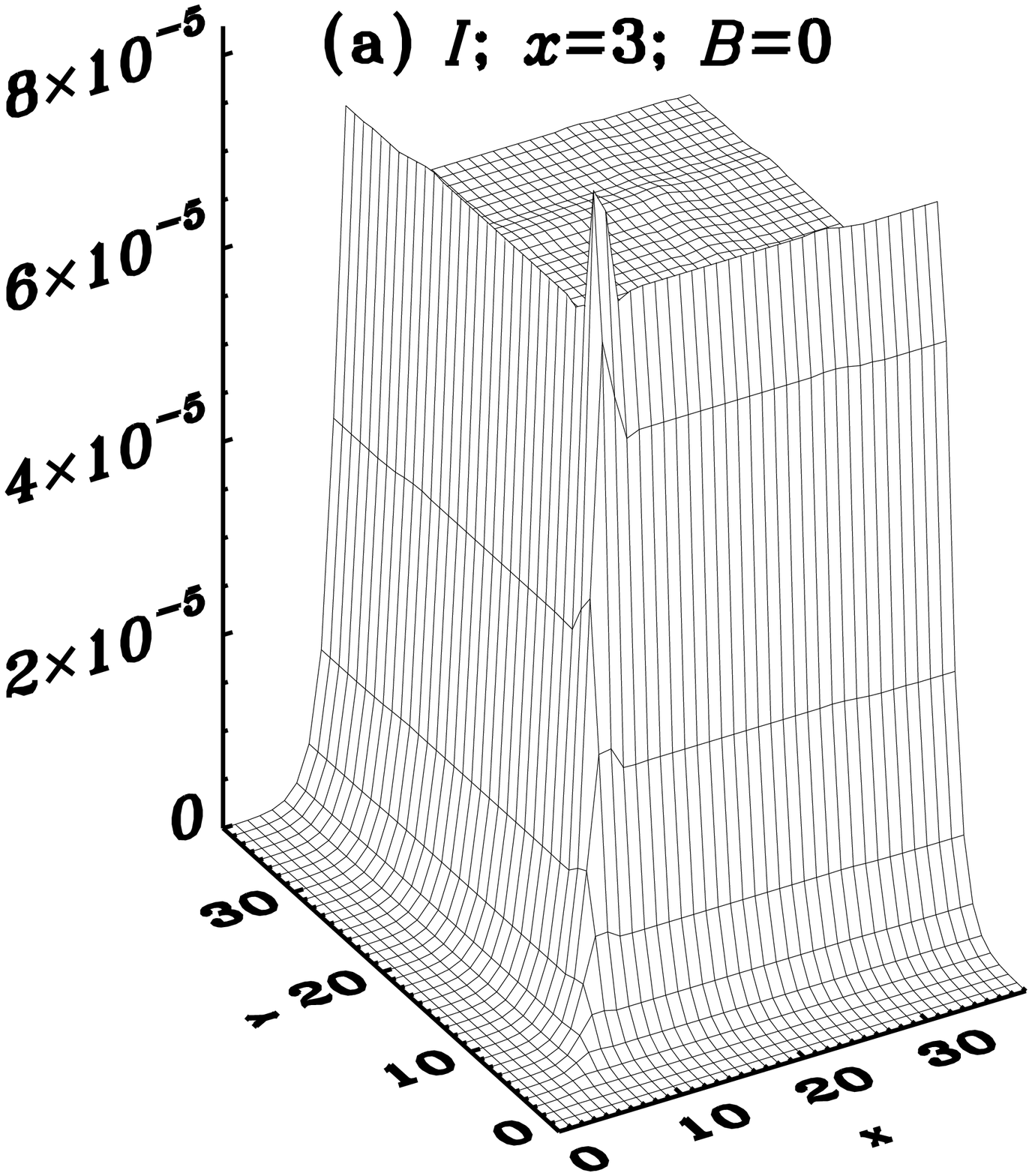}
\includegraphics[scale=0.25]{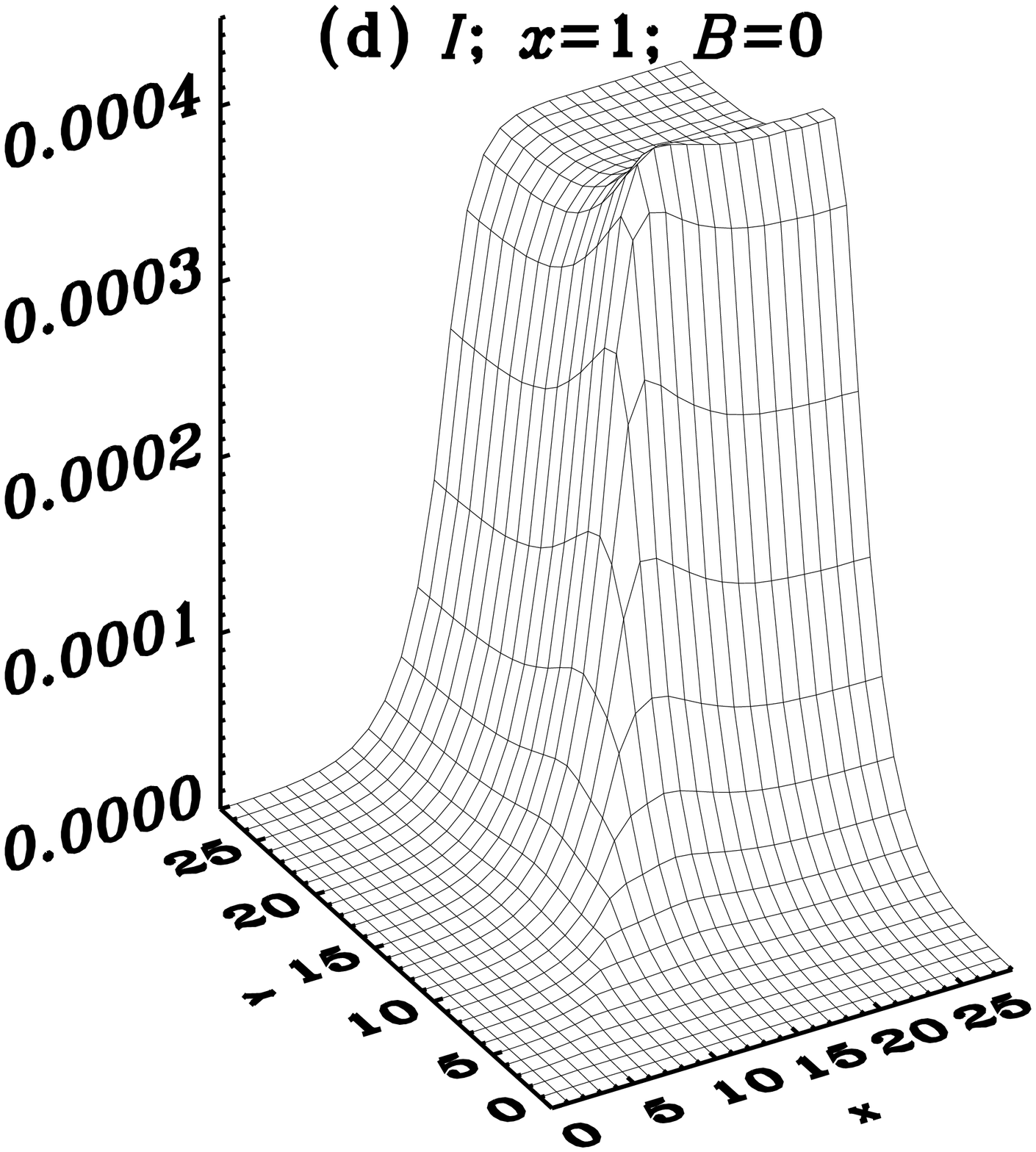}
\includegraphics[scale=0.25]{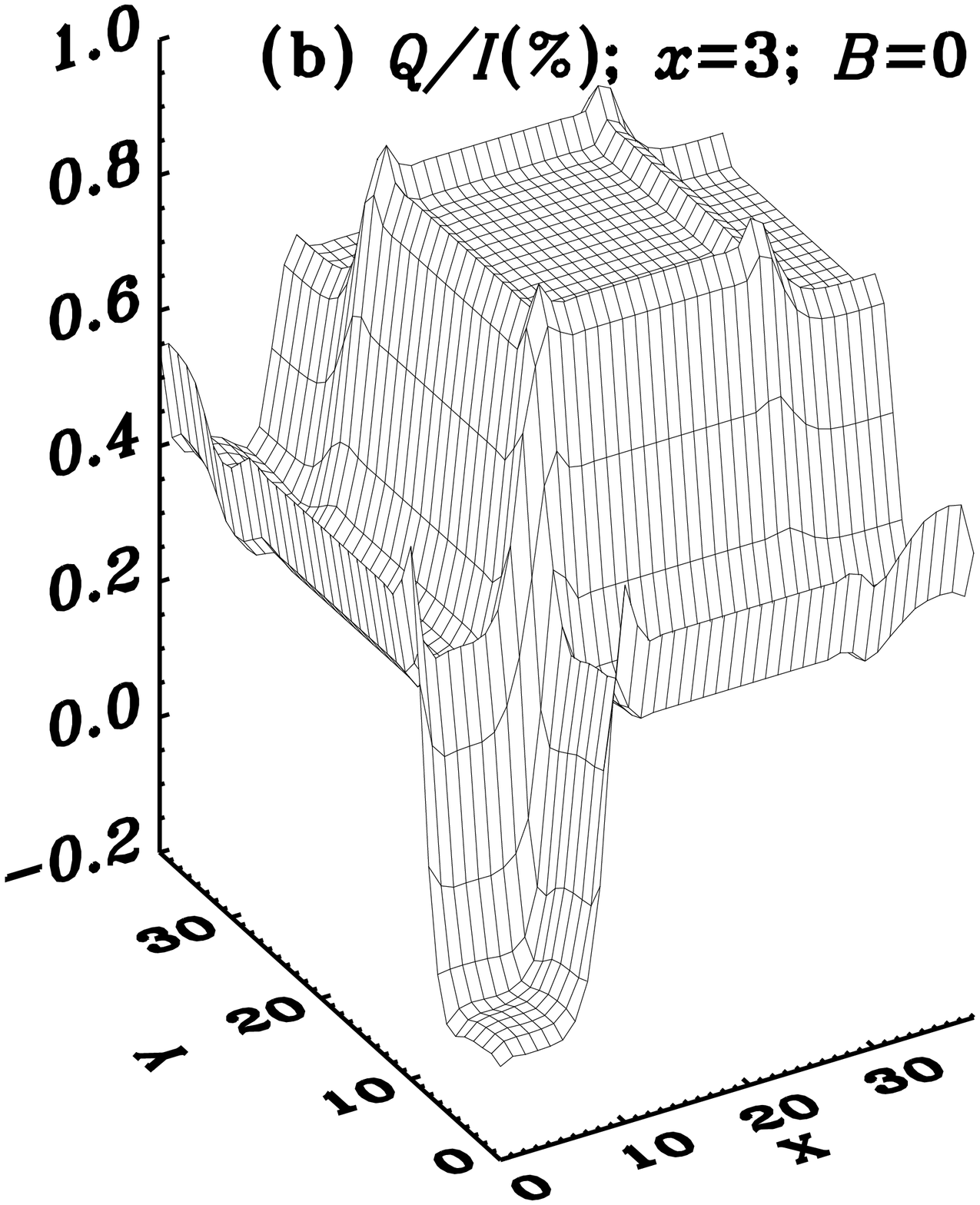}
\includegraphics[scale=0.25]{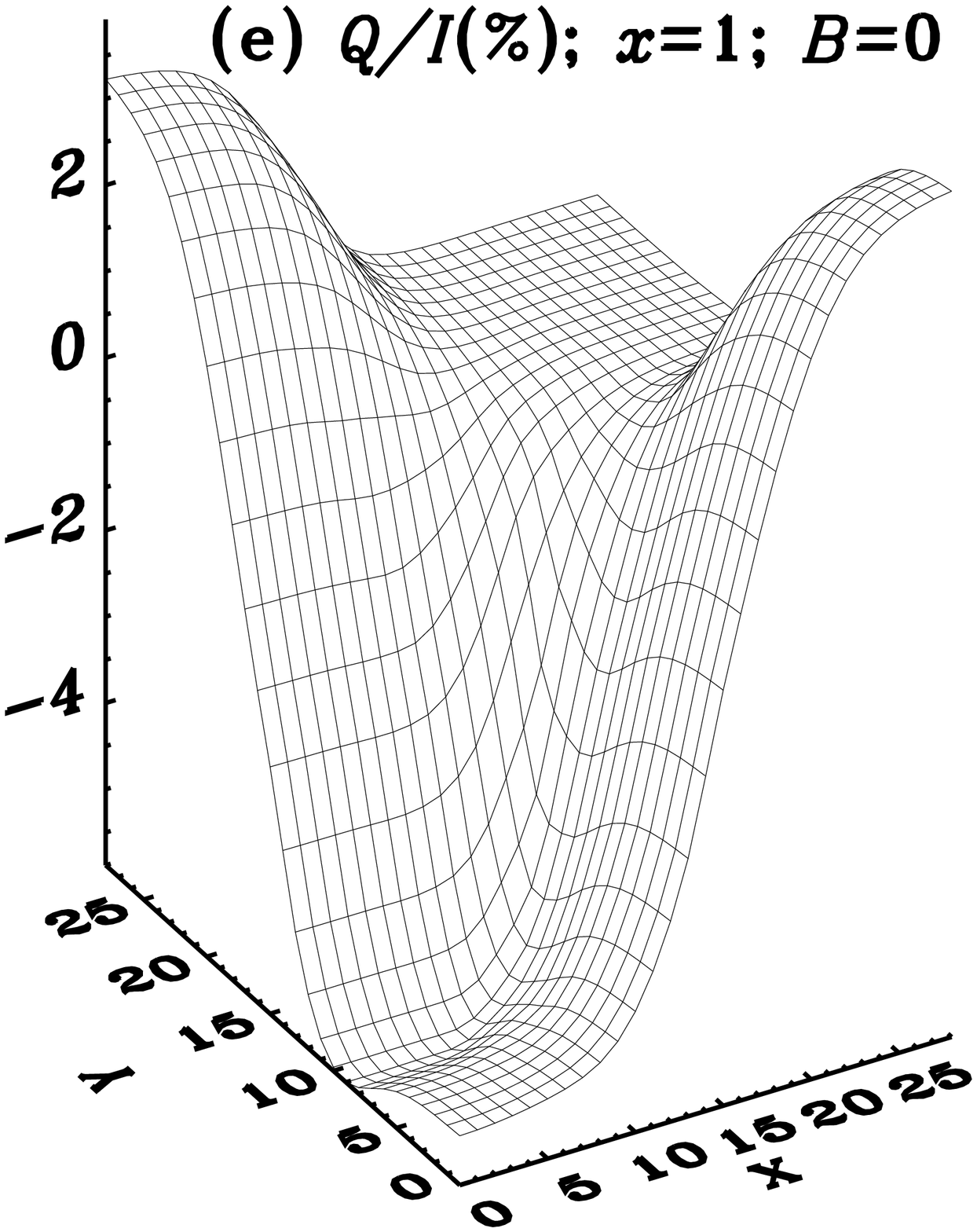}
\includegraphics[scale=0.25]{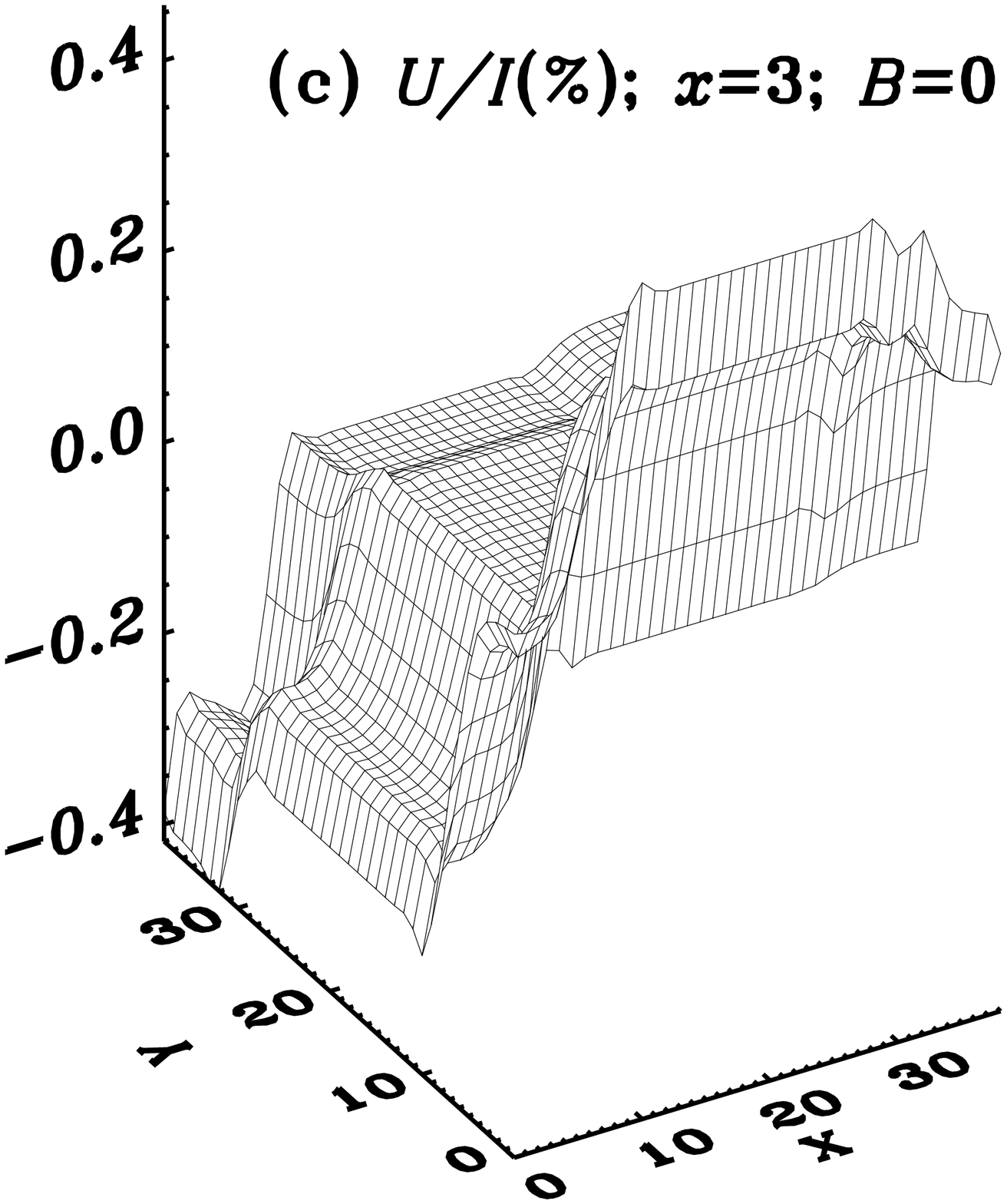}
\includegraphics[scale=0.25]{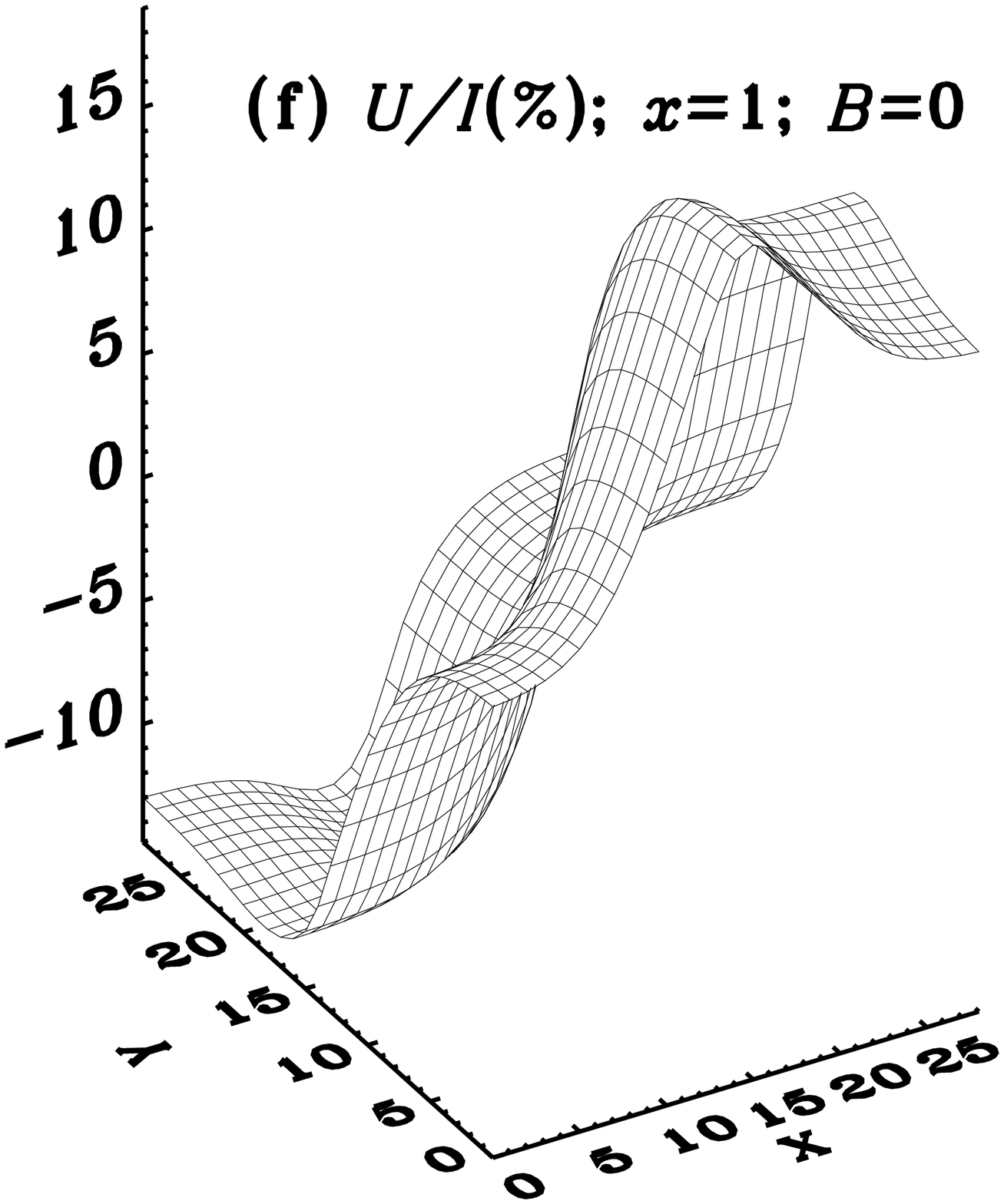}
\caption{The spatial distribution of ($Q/I, U/I$) on the
top surface of a 3D media. The $Q/I$ and $U/I$ are 
plotted as a function of the grid indices of $\tau_X$
and $\tau_Y$. 
The ray (viewing) direction is specified
by $(\mu,\varphi)$=$(0.11,60^{\circ})$. Left
panels represent a sheet structure and right panels
represent a rod structure when viewed along the $\pm Z$
direction.}
\label{fig-surface-thread}
\end{figure*}

In Figure~\ref{fig-surface-thread} we show spatial distribution of $I$,
$Q/I$ and $U/I$ on the top surface of two different kinds of 3D media. 
Here we have chosen $\bm{B}=0$ which is equivalent to the choice of 
a vertical magnetic field parallel to the $Z$ axis
(because, for this field geometry the Hanle effect goes to its non-magnetic
Rayleigh scattering limit). In view of the possible
applications, we consider a cuboid with $T_X=T_Y=2\times10^6$, 
$T_Z=20$ in the left panels (a, b, c) and a cuboid with 
$T_X=T_Y=20$, $T_Z=2\times10^6$ in the right panels (d, e, f). They represent
respectively a sheet and a rod like structure. 
For the chosen optical thickness configurations, the radiative transfer 
effects are mainly restricted to the line core ($x \le 3$) for the
ray emerging from the top surface. We show the results for $x=3$ (in the 
left panels) and $x=1$ (in the right panels), the frequencies for which the 
magnitudes of $Q/I$ and $U/I$ reach their maximum values.

In Figures~\ref{fig-surface-thread}(a) and (d) the intensities reach 
saturation values in the interiors of the top surface and drop to zero
at two of the visible boundaries (where a boundary condition of zero
intensity is imposed for our chosen ray emerging at the top surface).

In Figures~\ref{fig-surface-thread}(b) and (c) we see that $Q/I$ and $U/I$
take values $\le 1$\% everywhere on the top surface. The magnitude of $Q/I$
and $U/I$ for this case are relatively less than those for the semi-infinite 
3D atmospheres (compare with Figure~\ref{fig-surface}).
This can be understood using the following arguments.
We are showing the results for a ray with $(\mu,\varphi)=(0.11,60^{\circ})$ 
emerging from the top surface. 
The top surface for this figure refers to $\tau_Z=0$ where $\tau_Z$
is the optical depth measured inwards in the $Z$ direction. 
Using equations given in Appendix~\ref{appendixa} we can 
write approximate expressions for $Q$ and $U$ at the top surface as
\begin{eqnarray}
&&Q(\mu=0.11, \varphi=60^{\circ}, x)\approx\nonumber \\ 
&&\frac{-3}{2\sqrt{2}}
I^2_0(\mu=0.11, \varphi=60^{\circ}, x),
\label{qapprox}
\end{eqnarray}

\begin{eqnarray}
&&U(\mu=0.11, \varphi=60^{\circ}, x)\approx 
\nonumber \\
&&\frac{3}{2}I^{2,\rm x}_1(\mu=0.11, \varphi=60^{\circ}, x)
\nonumber \\
&&\!\!\!+\frac{\sqrt{3}}{2}
I^{2,\rm y}_1(\mu=0.11, \varphi=60^{\circ}, x).
\label{uapprox}
\end{eqnarray}
$I^2_0$ is controlled by the element $\Psi_{21}=3\cos^2\theta-1$ 
(see Appendix~\ref{appendixd}) which
appears in the scattering integral for $S^2_0$. The factor 
$\Psi_{21}=3\cos^2\theta-1$ represents the probability of scattering 
of photons incident from the direction $\theta$. For $\theta=0^{\circ}$ or
$\theta=180^{\circ}$ (vertical incidence) $\Psi_{21}$ is larger in magnitude
compared to the cases $\theta=90^{\circ}$ or $\theta=270^{\circ}$ (lateral
incidence). For $T_Z=20$
the medium is effectively optically thin (because $\epsilon T_Z<<1$)
in the $Z$ direction, and therefore photons easily escape in this direction.
Thus there are smaller number of photons for incidence along the vertical
direction when compared to the effectively thick case. 
For $T_Z=2\times10^6$ or $T_Z=2\times10^9$ the medium
is effectively optically thick (because $\epsilon T_Z>>1$) 
in the $Z$ direction and therefore
leaking of photons in this direction is reduced when compared
to the case of $T_Z=20$. In this way, for large values of $T_Z$ 
the probability of photons to be incident in the vertical direction 
is large. Therefore, as $T_Z$ increases 
the values of $I^2_0$ and hence $Q/I$ increase. 

For the chosen line of sight, Stokes $U$
is generated mainly by $I^{2,\rm x}_1$ and $I^{2,\rm y}_1$.
They are controlled by $\Psi_{31}$ and $\Psi_{41}$ elements
(see Appendix~\ref{appendixd}) both of which depend on the factor 
$\sin2\theta$. This implies that $\Psi_{31}$ and $\Psi_{41}$
are zero for both vertical and lateral incidence of photons.
These elements become larger when the incidence is predominant in the 
direction of $\theta=45^{\circ}$ or $\theta=135^{\circ}$. 
Using similar arguments as above we can understand the increase in the values
of $U/I$ with increasing values of $T_Z$.

The spatial distribution of $Q/I$ and $U/I$ is inhomogeneous in
both left and right panels for the chosen core frequencies, in 
contrast to the homogeneous distribution observed for
semi-infinite 3D atmospheres. The extent of inhomogenity is larger
for the left panels which correspond to smaller $T_Z$ value
than for the right panels. 
The spatial inhomogenity could also occur due to different optical
thicknesses along the 3 spatial directions leading to different
number of scatterings in the 3 directions (unlike the case of 
Figure~\ref{fig-surface} where $T_X=T_Y=T_Z$). In other words,
the inhomogeneities in $Q/I$ and $U/I$ can also be caused by a differential 
leaking of radiation in the $X$, $Y$ and $Z$ directions.

\section{CONCLUSIONS}
\label{conclusions}
This paper is dedicated to certain extensions of our previous
works (Paper I and Paper II) on polarized RT in multi-D media with PRD.

First, we present a generalization of the Stokes vector decomposition
technique developed in Paper I, to include the magnetic fields (Hanle effect).

Secondly, we generalize to the magnetic 3D RT, 
the efficient iterative method called the Pre-BiCG-STAB 
developed in Paper II for the non-magnetic 2D RT.

Thirdly, we use the more efficient 2D and 3D short characteristics
formal solutions, with appropriate generalizations to the 
present context. With the linear formal solver used
in Paper I, practically it is difficult to compute the solutions
in semi-infinite media. It is not the case with the short characteristics
former solution method. Indeed, the solutions presented in this
paper for the difficult case of semi-infinite media, prove this fact.

We present several benchmark solutions computed using the code, with 
all the above mentioned generalizations.
The main results of these solutions are the following.

The emergent $(Q/I, U/I)$ profiles in 1D media and the emergent,
spatially averaged $(Q/I, U/I)$ profiles in 2D and 3D media differ
significantly, both in non-magnetic and magnetic cases. The differences
are more pronounced in the wings of the $(Q/I, U/I)$ profiles.
The differences between the emergent, spatially averaged 
$(Q/I, U/I)$ profiles in 2D and 3D media are negligible in $Q/I$,
but noticeable in $U/I$.

In the non-magnetic case, at line center, the spatial distribution of
$Q/I$ and $U/I$ is homogeneous in the interior of the top surface,
but sharply raise near the edges. This is purely a multi-D
geometric effect. The presence of a magnetic field
modifies this distribution by causing a
depolarization (decrease in the magnitude) or re-polarization 
(increase in the magnitude) of $Q/I$ and $U/I$.
This is a natural consequence of the Hanle effect.
In the line wing frequencies, magnetic and non-magnetic 
spatial distributions look the same, as Hanle effect is confined 
to the line core. However, the spatial distribution 
in the line wing frequency is more inhomogeneous, and the
sharp raise of $Q/I$ and $U/I$ near the edges is more enhanced,
as compared to those at the line center.
This behavior at line wings is mainly due to the PRD effects.
These characteristics are not noticeable if the CRD assumption
is used in line formation studies.

We have developed efficient techniques to solve polarized RT
in multi-D media with PRD as the scattering mechanism. In future, we 
try to apply these methods to understand the linear polarization
observed in the spatially resolved structures on the Sun.
\acknowledgments
\noindent
We would like to thank 
Prof. H. Frisch for useful suggestions which helped to 
improve the manuscript. We thank Dr. Sampoorna for useful
discussions.

\appendix
\section{EXPANSION OF STOKES PARAMETERS INTO THE IRREDUCIBLE
COMPONENTS}
\label{appendixa}
The Stokes parameters and the irreducible Stokes vector are related
through the following expressions. They are already given in
\citet{hf07}. However we present these expressions here
for an easy reference.
\begin{eqnarray}
&&I(\bm{r}, \bm{\Omega}, x) = I^0_0 +
\frac{1}{2 \sqrt{2}} (3 \cos^2\theta -1) I^2_0 \nonumber \\
&&-\sqrt{3} \cos \theta \sin \theta (I^{2, {\rm x}}_1 
\cos \varphi-I^{2, {\rm y}}_1 \sin \varphi) \nonumber \\ 
&&+ \frac{\sqrt{3}}{2} (1-\cos^2\theta)
(I^{2, {\rm x}}_2 \cos 2 \varphi-I^{2, {\rm y}}_2 \sin 2 \varphi), \nonumber \\
\label{transform-1}
\end{eqnarray}

\begin{eqnarray}
&&Q(\bm{r}, \bm{\Omega}, x)= -\frac{3}{2 \sqrt{2}} 
(1- \cos^2\theta) I^2_0 \nonumber \\
&&-\sqrt{3} \cos \theta \sin \theta (I^{2, {\rm x}}_1 
\cos \varphi-I^{2, {\rm y}}_1 \sin \varphi) \nonumber \\ 
&&-\frac{\sqrt{3}}{2} (1+\cos^2\theta)
(I^{2, {\rm x}}_2 \cos 2 \varphi-I^{2, {\rm y}}_2 \sin 2 \varphi),\nonumber \\
\label{transform-2}
\end{eqnarray}

\begin{eqnarray}
&&U(\bm{r}, \bm{\Omega}, x) = \sqrt{3} \sin \theta
(I^{2, {\rm x}}_1 \sin \varphi+I^{2, {\rm y}}_1 \cos \varphi) \nonumber \\ 
&&+ \sqrt{3} \cos \theta 
(I^{2, {\rm x}}_2 \sin 2 \varphi+I^{2, {\rm y}}_2 \cos 2 \varphi).
\label{transform-3}
\end{eqnarray}

The irreducible components in the above equations also depend on
$\bm{r}$, $\bm{\Omega}$, $x$ and $\bm{B}$.
\section{THE REDISTRIBUTION MATRICES IN THE IRREDUCIBLE TENSORIAL FORM}
\label{appendixb}
In this paper we use the redistribution matrices defined 
under the approximation level III of \citet{bom97b}.
The expressions listed below are already given in \citet{bom97b}. 
We give them here for the sake of completeness. 
The branching ratios \citep[see][]{bom97b} are given by
\begin{equation}
\alpha=\frac{\Gamma_R}{\Gamma_R+\Gamma_E+\Gamma_I},
\label{alp}
\end{equation}

\begin{equation}
\beta^{(K)}=\frac{\Gamma_R}{\Gamma_R+D^{(K)}+\Gamma_I},
\label{beta0}
\end{equation}
with $D^{(0)}=0$ and $D^{(2)}=c \Gamma_E$, where $c$ is a constant,
taken to be 0.379 \citep[see][]{mf92}.

The Hanle $\Gamma_B$ coefficient \citep[see][]{bom97b} takes 
two different forms, namely
\begin{equation}
\Gamma_B=\Gamma'_K=\beta^{(K)}\Gamma,\quad\Gamma_B=\Gamma''=\alpha\Gamma,
\label{gamma}
\end{equation}
with
\begin{equation}
\Gamma=g_J\, \frac{2\pi eB}{2m_e\Gamma_R}
\end{equation}
where ${eB}/{2m_e}$ is the Larmor frequency
of the electron in the magnetic field (with $e$ and $m_e$ being
the charge and mass of the electron). Here $B$ is the magnetic field
strength. 
The expressions for the redistribution matrices 
given in \citet{bom97b} involve a cut-off frequency
$v_c(a)$, which is given by the solution of the equation
\begin{equation}
\frac{1}{\sqrt{\pi}} e^{-v^2}=\frac{a}{\pi}\frac{1}{v^2+a^2}, 
\end{equation}
and a constant $z=2\sqrt{2}+2$ coming from the angle-averaging
process.

\noindent
If
\begin{flalign}
\noindent & zv_c(a) |x'| - (x^2+x'^2) < (z-1) v_c^2(a) \quad
\textrm{and} &\nonumber \\ 
\noindent & z v_c(a) |x| - (x^2+x'^2) < (z-1) v_c^2(a) \quad
\textrm{and} &\nonumber\\
\noindent & |x'| < \sqrt{2}v_c(a) \quad\textrm{and} \quad |x| <
\sqrt{2}v_c(a),& 
\end{flalign}
\noindent
then domain 1\,:

\begin{eqnarray}
&&\!\!\!\!\!\!
P^K_{Q, \rm III}(j,\bm{\Omega}',\bm{B})=
\sum_{Q'}\Big\{\beta^{(K)} 
\mathcal{M}^K_{QQ'}(\bm{B};\Gamma'_K)\nonumber \\
&&\!\!\!\!\!\!
-\alpha \mathcal{M}^K_{QQ'}(\bm{B};\Gamma'')\Big\} (-1)^{Q'}
\mathcal{T}^K_{-Q'}(j, \bm{\Omega}'),
\nonumber\\
&&\!\!\!\!\!\!=\sum_{Q'}\overline{\mathcal{M}}^{K(1)}_{QQ', \rm III}(\bm{B})
(\mathcal{T}^K_{Q'})^{*}(j, \bm{\Omega}').
\label{pkq1}
\end{eqnarray}

\noindent
elseif
\begin{flalign}
\noindent &|x'| < v_c(a) \quad\textrm{or}\quad |x| < v_c(a),&
\end{flalign}

\noindent
then domain 2\,:

\begin{eqnarray}
&&\!\!\!\!\!\!
P^K_{Q, \rm III}(j,\bm{\Omega}',\bm{B})=
[\beta^{(K)}-\alpha]
\nonumber \\
&&\!\!\!\!\!\!
\times \sum_{Q'}\mathcal{M}^K_{QQ'}(\bm{B};\Gamma'_K)
(-1)^{Q'}\mathcal{T}^K_{-Q'}(j, \bm{\Omega}'),
\nonumber\\
&&\!\!\!\!\!\!=\sum_{Q'}\overline{\mathcal{M}}^{K(2)}_{QQ', \rm III}(\bm{B})
(\mathcal{T}^K_{Q'})^{*}(j, \bm{\Omega}').
\label{pkq2}
\end{eqnarray}

\noindent
else domain 3\,:
\begin{eqnarray}
&&\!\!\!\!\!\!
P^K_{Q, \rm III}(j,\bm{\Omega}',\bm{B})=
[1-\alpha/\beta^{(K)}]\Big\{[\beta^{(K)}-\alpha]
\nonumber \\
&&\!\!\!\!\!\!
\times \sum_{Q'}\mathcal{M}^K_{QQ'}(\bm{B};\Gamma'_K)
(-1)^{Q'}\mathcal{T}^K_{-Q'}(j, \bm{\Omega}')
\nonumber \\
&&\!\!\!\!\!\!
+\alpha\sum_{Q'}(-1)^{Q'}\mathcal{T}^K_{-Q'}(j, \bm{\Omega}')\Big\},
\nonumber\\
&&\!\!\!\!\!\!=\sum_{Q'}\overline{\mathcal{M}}^{K(3)}_{QQ', \rm III}(\bm{B})
(\mathcal{T}^K_{Q'})^{*}(j, \bm{\Omega}').
\label{pkq3}
\end{eqnarray}
endif.
\noindent
If
\begin{flalign}
\noindent &x(x+x') < 2 v_c^2(a)\quad \textrm{and} \quad 
x'(x+x') < 2 v_c^2(a),&
\end{flalign}

\noindent
then domain 4\,:
\begin{eqnarray}
&&\!\!\!\!\!\!
P^K_{Q, \rm II}(j,\bm{\Omega}',\bm{B})\nonumber \\
&&\!\!\!\!\!\!=
\alpha\sum_{Q'}\mathcal{M}^K_{QQ'}(\bm{B};\Gamma'')
(-1)^{Q'}\mathcal{T}^K_{-Q'}(j, \bm{\Omega}').
\nonumber\\
&&\!\!\!\!\!\!=\sum_{Q'}\overline{\mathcal{M}}^{K(4)}_{QQ', \rm II}(\bm{B})
(\mathcal{T}^K_{Q'})^{*}(j, \bm{\Omega}'),
\label{pkq4}
\end{eqnarray}

\noindent
else domain 5\,:
\begin{eqnarray}
&&\!\!\!\!\!\!P^K_{Q, \rm II}(j,\bm{\Omega}',\bm{B})=
\alpha\sum_{Q'}(-1)^{Q'}\mathcal{T}^K_{-Q'}(j, \bm{\Omega}'),
\nonumber\\
&&\!\!\!\!\!\!=\sum_{Q'}\overline{\mathcal{M}}^{K(5)}_{QQ', \rm II}(\bm{B})
(\mathcal{T}^K_{Q'})^{*}(j, \bm{\Omega}').
\label{pkq5}
\end{eqnarray}
\noindent
endif.\\

The symbols $\overline{\mathcal{M}}^{K(i)}_{QQ', \rm II,III}(\bm{B})$,
$i=1, 2, 3, 4, 5$ have different expressions in different 
frequency domains. They implicitly 
contain the respective branching ratios and the Hanle $\Gamma$ parameter
depending upon the domain. 
\section{THE REDISTRIBUTION MATRICES IN THE MATRIX FORM}
\label{appendixc}
We introduce the diagonal matrices
\begin{equation}
\hat{\alpha}=\alpha \hat E,
\label{alpha}
\end{equation}
with $\hat E$ the identity matrix,
\begin{equation}
\hat{\beta}=\textrm{diag}
\{\beta^{(0)},\beta^{(2)},\beta^{(2)},\beta^{(2)},\beta^{(2)},\beta^{(2)}\},
\label{beta}
\end{equation}
\begin{eqnarray}
&&{\hat{\mathcal{ F}}}=\textrm{diag}
\Bigg\{1-{\alpha\over\beta^{(0)}},1-{\alpha\over\beta^{(2)}},
1-{\alpha\over\beta^{(2)}},\nonumber \\
&&1-{\alpha\over\beta^{(2)}},
1-{\alpha\over\beta^{(2)}},1-{\alpha\over\beta^{(2)}}\Bigg\}.
\label{F-alphaandbeta}
\end{eqnarray}
The real matrices $\hat{M}^{(i)}_{\rm II}(\bm{B})$ and 
$\hat{M}^{(i)}_{\rm III}(\bm{B})$ have following expressions in
different domains.\\

\noindent
In domain 1:\,\\

\begin{eqnarray}
\hat{M}^{(1)}_{\rm III}(\bm{B})=
\Bigg\{\hat{\beta}\hat{M}(\bm{B},\Gamma'_2)
-\hat{\alpha}\hat{M}(\bm{B},\Gamma'')\Bigg\}.
\label{mr-dom-1}
\end{eqnarray}

\noindent
In domain 2:\,\\

\begin{eqnarray}
\hat{M}^{(2)}_{\rm III}(\bm{B})=\Bigg\{\Big[\hat{\beta}-\hat{\alpha}\Big]
\hat{M}(\bm{B},\Gamma'_K)\Bigg\}.
\label{mr-dom-2}
\end{eqnarray}

\noindent
In domain 3:\,\\
\begin{eqnarray}
\hat{M}^{(3)}_{\rm III}(\bm{B})={\hat{\mathcal{F}}}
\Bigg\{\Big[\hat{\beta}-\hat{\alpha}\Big] \hat{M}(\bm{B},\Gamma'_2)
+\hat{\alpha}\Bigg\}.
\label{mr-dom-3}
\end{eqnarray}

\noindent
In domain 4:\,\\
\begin{eqnarray}
\hat{M}^{(4)}_{\rm II}(\bm{B})=\hat{\alpha} \hat{M}(\bm{B},\Gamma'').
\label{mr-dom-4}
\end{eqnarray}

\noindent
In domain 5:\,\\
\begin{eqnarray}
\hat{M}^{(5)}_{\rm II}(\bm{B})=\hat{\alpha}.
\label{mr-dom-5}
\end{eqnarray}
\section{The scattering phase matrix in real form in the reduced basis}
\label{appendixd}
The elements of the matrix $\hat{\Psi}$ are already given in Appendix A of 
Paper I. However we have found that there were some typographical errors there.
We give here the elements again, correcting those typographical errors.
\begin{equation}
\hat{\Psi}^r = 
\left ( \begin{array}{cccccc}
\vspace{.1cm}
{\Psi}_{11}&{\Psi}_{12}&{\Psi}_{13}&{\Psi}_{14}
&{\Psi}_{15}&{\Psi}_{16} \\
\vspace{.3cm}
{\Psi}_{12}&{\Psi}_{22}&{\Psi}_{23}&{\Psi}_{24}
&{\Psi}_{25}&{\Psi}_{26} \\
\vspace{.3cm}
\frac{1}{2}{\Psi}_{13}&\frac{1}{2}{\Psi}_{23}&{\Psi}_{33}
&{\Psi}_{34}&{\Psi}_{35}&{\Psi}_{36} \\
\vspace{.3cm}
\frac{1}{2}{\Psi}_{14}&\frac{1}{2}{\Psi}_{24}&{\Psi}_{34}
&{\Psi}_{44}&{\Psi}_{45}&{\Psi}_{46} \\
\vspace{.3cm}
\frac{1}{2}{\Psi}_{15}&\frac{1}{2}{\Psi}_{25}&{\Psi}_{35}
&{\Psi}_{45}&{\Psi}_{55}&{\Psi}_{56} \\
\vspace{.3cm}
\frac{1}{2}{\Psi}_{16}&\frac{1}{2}{\Psi}_{26}&{\Psi}_{36}
&{\Psi}_{46}&{\Psi}_{56}&{\Psi}_{66} \\
\end{array} \right),
\label{psi-matrix-symmetry}
\end{equation}
where the distinct matrix elements are:
\begin{eqnarray}
&&{\Psi}_{11}=1 
;\quad {\Psi}_{12}=\frac{1}{2 \sqrt 2} (3 \cos^2 \theta -1);\nonumber \\
&&{\Psi}_{13}=-\frac{\sqrt 3}{2} \sin 2 \theta \cos \varphi
;\quad{\Psi}_{14}= \frac{\sqrt 3}{2} \sin 2 \theta \sin \varphi
;\quad{\Psi}_{15}= \frac{\sqrt 3}{2} \sin^2 \theta \cos 2 \varphi;
\nonumber\\
&&{\Psi}_{16}= -\frac{\sqrt 3}{2} \sin^2 \theta \sin 2 \varphi;
\quad{\Psi}_{22}=\frac{1}{4} (9 \cos^4 \theta - 12 \cos^2 \theta 
+ 5);\nonumber \\
&&{\Psi}_{23}=  \frac{\sqrt 3}{4 \sqrt 2} \sin 2 \theta 
(1-3 \cos 2 \theta) \cos \varphi
;\quad{\Psi}_{24}= -\frac{\sqrt 3}{4 \sqrt 2} \sin 2 \theta 
(1-3 \cos 2 \theta) \sin \varphi;
\nonumber \\
&&{\Psi}_{25}=\frac{\sqrt 3}{2 \sqrt 2} \sin^2 \theta
(1+3\cos^2 \theta) \cos 2 \varphi
;\quad{\Psi}_{26}= -\frac{\sqrt 3}{2 \sqrt 2}\sin^2 \theta
(1+3\cos^2 \theta) \sin 2 \varphi;
\nonumber \\
&&{\Psi}_{33}= \frac{3}{4} \sin^2 \theta [(1+2 \cos^2 \theta)-
(1-2 \cos^2 \theta) \cos 2 \varphi ];
\nonumber \\
&&{\Psi}_{34}=\frac{3}{4} \sin^2 \theta (1-2 \cos^2 \theta) 
\sin 2 \varphi
;\quad{\Psi}_{35}=\frac{3}{16} \sin 2 \theta [(3+\cos 2 \theta) 
\cos \varphi- (1-\cos 2 \theta) \cos 3 \varphi];
\nonumber \\
&&{\Psi}_{36}= -\frac{3}{16} \sin 2 \theta 
[ (3+\cos 2 \theta) \sin \varphi - (1-\cos 2 \theta) \sin 3 \varphi];
\nonumber \\
&&{\Psi}_{44}=\frac{3}{4} \sin^2 \theta [(1+2 \cos^2 \theta)
+(1-2 \cos^2 \theta) \cos 2 \varphi];
\nonumber \\
&&{\Psi}_{45}=\frac{3}{16} \sin 2 \theta [(3+\cos 2 \theta) 
\sin \varphi+(1-\cos 2 \theta) \sin 3 \varphi];
\nonumber \\
&&{\Psi}_{46}=\frac{3}{16} \sin 2 \theta [(3+\cos 2 \theta) 
\cos \varphi+(1-\cos 2 \theta) \cos 3 \varphi];
\nonumber \\
&&{\Psi}_{55}=\frac{3}{16}[(1+6 \cos^2 \theta+\sin^4\theta 
+\cos^4\theta)+(1-2 \cos^2 \theta +\cos^4 \theta+\sin^4 \theta) 
\cos 4 \varphi];
\nonumber \\
&&{\Psi}_{56}=-\frac{3}{16}[(1-2\cos^2 \theta
+\cos^4\theta+\sin^4\theta)\sin 4 \varphi ];
\nonumber \\
\!\!\!\!\!\!\!&&{\Psi}_{66}=\frac{3}{16}
[(1+6 \cos^2 \theta+\sin^4\theta 
+\cos^4\theta)-(1-2 \cos^2 \theta +\cos^4 \theta
+\sin^4 \theta) \cos 4 \varphi].
\label{phi-matrix-elements}
\end{eqnarray}
The elements of the matrix $\hat{\Psi}$
satisfy certain symmetry properties with respect to the main diagonal.
Hence the number of independent elements are only 21.

%%%%%%%%%%%%%%%%%%%%%%%%%%%%%%%%%%%%%%%%%%%%%%%%%%%%%%%%%%%
\begin{table*}
\begin{center}
\caption{The 12-point Carlsson type B quadrature for the 
azimuth angle $\varphi$. The corresponding values of 
$\sin \varphi$, $\cos \varphi$, $\sin 2\varphi$ and $\cos 2\varphi$ are
given for the purpose of discussion.}
\vspace{0.6cm}
\label{table_1}
\begin{tabular}{crrrr}
\tableline\tableline
$\varphi_i$ (in degrees)&$\sin \varphi$&$\cos \varphi$&$\sin 2\varphi$&$\cos 2\varphi$\\
\tableline\tableline
30&0.5&0.866&0.866&0.5\\
\tableline
45&0.707&0.707&1&0\\
\tableline
60&0.866&0.5&0.866&-0.5\\
\tableline
120&0.866&-0.5&-0.866&-0.5\\
\tableline
135&0.707&-0.707&-1&0\\
\tableline
150&0.5&-0.866&-0.866&0.5\\
\tableline
210&-0.5&-0.866&0.866&0.5\\
\tableline
225&-0.707&-0.707&1&0\\
\tableline
240&-0.866&-0.5&0.866&-0.5\\
\tableline
300&-0.866&0.5&-0.866&-0.5\\
\tableline
315&-0.707&0.707&-1&0\\
\tableline
330&-0.5&0.866&-0.866&0.5\\
\tableline
\end{tabular}
\end{center}
\end{table*}
%%%%%%%%%%%%%%%%%%%%%%%%%%%%%%%%%%%%%%%%%%%%%%%%%%%%%%%%%%%
%%%%%%%%%%%%%%%%%%%%%%%%%%%%%%%%%%%%%
%%%%%%%%%%%%%%%%%%%%%%%%%%%%%%%%%%%%%%%%%%%
%%%%%%%%%%%%%%%%%%%%%%%%%%%%%%%%%%%%% 
%%%%%%%%%%%%%%%%%%%%%%%%%%%%%%%%%%%%%%%%%%%%%%%
%%%%%%%%%%%%%%%%%%%%%%%%%%%%%%%%%%%%%
%%%%%%%%%%%%%%%%%%%%%%%%%%%%%%%%%%%%%%%%%%%%%%%
%%%%%%%%%%%%%%%%%%%%%%%%%%%%%%%%%%%%%
%%%%%%%%%%%%%%%%%%%%%%%%%%%%%%%%%%%%%%%%%%%%%%%
%%%%%%%%%%%%%%%%%%%%%%%%%%%%%%%%%%%%%
%%%%%%%%%%%%%%%%%%%%%%%%%%%%%%%%%%%%%%%%%%%%%%%
%%%%%%%%%%%%%%%%%%%%%%%%%%%%%%%%%%%%%
%%%%%%%%%%%%%%%%%%%%%%%%%%%%%%%%%%%%
%%%%%%%%%%%%%%%%%%%%%%%%%%%%%%%%%%%%
%%%%%%%%%%%%%%%%%%%%%%%%%%%%%%%%%%%%%%%%%%%%%%%
%%%%%%%%%%%%%%%%%%%%%%%%%%%%%%%%%%%%%
%%%%%%%%%%%%%%%%%%%%%%%%%%%%%%%%%%%%%%%%%%%%%%%
%%%%%%%%%%%%%%%%%%%%%%%%%%%%%%%%%%%%%
%%%%%%%%%%%%%%%%%%%%%%%%%%%%%%%%%%%%%%%%%%%%%%%
%%%%%%%%%%%%%%%%%%%%%%%%%%%%%%%%%%%%%
%%%%%%%%%%%%%%%%%%%%%%%%%%%%%%%%%%%%%%%%%%%%%%%
%%%%%%%%%%%%%%%%%%%%%%%%%%%%%%%%%%%%%
%%%%%%%%%%%%%%%%%%%%%%%%%%%%%%%%%%%%%%%%%%%%%%%
%%%%%%%%%%%%%%%%%%%%%%%%%%%%%%%%%%%%%
%%%%%%%%%%%%%%%%%%%%%%%%%%%%%%%%%%%%%
%%%%%%%%%%%%%%%%%%%%%%%%%%%%%%%%%%%%%
%%%%%%%%%%%%%%%%%%%%%%%%%%%%%%%%%%%%%%%%%%%%%%%
%%%%%%%%%%%%%%%%%%%%%%%%%%%%%%%%%%%%%
%%%%%%%%%%%%%%%%%%%%%%%%%%%%%%%%%%%%%
\end{document}